\documentclass[twocolumn,prl,amsmath,amssymb]{revtex4-1}
\usepackage[latin9]{inputenc}
\usepackage{amsmath}
\usepackage{amssymb}
\usepackage{graphicx}
\usepackage[unicode=true,pdfusetitle,
 bookmarks=true,bookmarksnumbered=false,bookmarksopen=false,
 breaklinks=false,pdfborder={0 0 1},backref=false,colorlinks=false]
 {hyperref}
\begin{document}
\title{Finite temperature mean-field theory with intrinsic non-hermitian
structures for Bose gases in optical lattices }
\author{Liang He}
\email{liang.he@scnu.edu.cn}

\affiliation{Guangdong Provincial Key Laboratory of Quantum Engineering and Quantum
Materials, SPTE, South China Normal University, Guangzhou 510006,
China}
\author{Su Yi}
\email{syi@itp.ac.cn}

\affiliation{CAS Key Laboratory of Theoretical Physics, Institute of Theoretical
Physics, Chinese Academy of Sciences, Beijing 100190, China}
\affiliation{School of Physical Sciences \& CAS Center for Excellence in Topological
Quantum Computation, University of Chinese Academy of Sciences, Beijing
100049, China}
\begin{abstract}
We reveal a divergent issue associated with the mean-field theory
for Bose gases in optical lattices constructed by the widely used
straightforward mean-field decoupling of the hopping term, where the
corresponding mean-field Hamiltonian generally assumes no lower energy
bound once the spatial dependence of the mean-field superfluid order
parameter is taken into account. Via a systematic functional integral
approach, we solve this issue by establishing a general finite temperature
mean-field theory that can treat any possible spatial dependence of
the order parameter without causing the divergent issue. Interestingly,
we find the theory generally assumes an intrinsic non-hermitian structure
that originates from the indefiniteness of the hopping matrix of the
system. Within this theory, we develop an efficient approach for investigating
the physics of the system at finite temperature, where properties
of the system can be calculated via straightforward investigation
on the saddle points of an effective potential function for the order
parameter. We illustrate our approach by investigating the finite
temperature superfluid transition of Bose gases in optical lattices.
Since the underlying finite temperature mean-field theory is quite
general, this approach can be straightforwardly applied to investigate
the finite temperature properties of related systems with phases possessing
complex spatial structures. 
\end{abstract}
\maketitle

\section{Introduction}

Since the first experimental realization of the Bose-Hubbard model
\citep{Fisher_PRB_1989} with ultracold atoms in optical lattices
\citep{Greiner_Nature_2002}, ultracold gases in optical lattices
have become one of the most important experimental platforms \citep{Bloch_RMP_2008}
to investigate rich physics that are relevant for a wide range of
different fields, ranging from solid-state physics, over condensed
matter physics, to high energy physics. This is attributed to the
tremendous experimental development in ultracold gases in optical
lattices in the last two decades, including realizing, for instance,
optical lattices with different dimensionality and geometries \citep{Stoferle_PRL_2004,Spielman_PRL_2007,Becker_NJP_2010,Jo_PRL_2012},
different types of interactions \citep{Moses_Science_2015,Baier_Science_2016,Landig_Nature_2016},
artificial gauge fields \citep{Aidelsburger_PRL_2011,Duca_Science_2015,Wu_Science_2016},
etc, which at the same time gives rise to considerable types of systems
theat support rich unconventional quantum phases with nontrivial spatial
structures. 

A case in point is ultracold Bose gases in optical lattices, where
for this type of systems with, for instance, long-range interactions
\citep{Moses_Science_2015,Baier_Science_2016,Landig_Nature_2016},
artificial gauge fields \citep{Cooper_PRL_2011,Aidelsburger_PRL_2011,Duca_Science_2015,Wu_Science_2016},
etc, experimental and theoretical investigations \citep{Aidelsburger_PRL_2011,Duca_Science_2015,Landig_Nature_2016,Yi_PRL_2007,Li_PRA_2013,He_PRA_2021,Cole_PRL_2012,Radic_PRL_2012,Cai_PRA_2012,Hickey_PRL_2014,He_PRA_2015}
have shown that they can support exotic phases with nontrivial spatial
structures, such as charge-density waves, spin-density waves, unconventional
superfluid, etc. In investigations of these systems, straightforward
mean-field decoupling approach \citep{Oosten_PRA_2001,Sachdev_QPT_2011}
and bosonic Gutzwiller mean-field variational approach \citep{Jaksch_PRL_1998,Krauth_PRB_1992}
are two of the most widely employed theoretical tools that have efficiently
revealed considerable nontrivial physics of these systems at zero
temperature. However, concerning finite temperature properties of
these systems, which are naturally indispensable for relevant experiments
and the thorough understanding of their physical behavior, these two
approaches seem not as efficient to be employed as in the zero temperature
investigations \citep{Hickey_PRL_2014}, not even to mention that
the legitimacy of the straightforward mean-field decoupling is not
self-evident in different application scenarios and often overlooked
in the literature \citep{Fisher_PRB_1989,Oosten_PRA_2001,Sachdev_QPT_2011}.
This thus poses the natural demand for a reliable mean-field theory
that could efficiently investigate the finite temperature properties
of these systems.

We address this demand for the type of mean-field theory where the
mean-field treatment is performed on the hopping term of the system
to investigate the superfluid transition. Using single-component Bose
gases in optical lattices as the concrete type of systems, we show
that the widely used straightforward mean-field decoupling of the
hopping term \citep{Oosten_PRA_2001,Sachdev_QPT_2011} generally results
in problematic mean-field Hamiltonians with no lower energy bounds
once the spatial dependence of the mean-field superfluid order parameter
is taken into account (cf.~Eq.~(\ref{eq:H_SMFD_two_sub_lat}) and
the discussion below). We solve this problem by establishing the proper
finite temperature mean-field theory {[}cf.~Eq.~(\ref{eq:Z_MF}){]}
via a systematic functional integral approach. In particular, we find
the proper mean-field Hamiltonian generally assumes an intrinsic non-hermitian
structure {[}cf.~Eqs.~(\ref{eq:HMF_general_form}, \ref{eq:HSS_general_form}){]}
that originates from the indefiniteness of the hopping matrix of the
system. Based on this non-hermitian mean-field Hamiltonian, we develop
an efficient and versatile approach for investigating the physics
of the system at finite temperatures, where properties of the system
can be calculated via straightforward investigations on the saddle
points of an effective potential function for the order parameter
{[}cf.~Eqs.~(\ref{eq:Z_MF_Omega}, \ref{eq:Omega_general}) and
Fig.~\ref{Fig_1_Finite_T_Phase_diagram}{]}. We illustrated our approach
by using both a homogeneous and an inhomogeneous two-sublattice finite
temperature mean-field theory to investigate the finite temperature
superfluid transition of Bose gases in optical lattices, the results
of which at low temperature agree with the well-established results
from, for instance, bosonic Gutzwiller variational ansatz \citep{Jaksch_PRL_1998,Krauth_PRB_1992}.
And we directly map out the finite temperature phase diagram of the
system (cf.~Fig.~\ref{Fig_1_Finite_T_Phase_diagram}) and show how
Mott lobes ``melt'' in the presence of thermal fluctuations (cf.~Fig.~\ref{Fig_2_Mott_lobes_at_different_T_Hom_MFT}
and Fig.~\ref{Fig_3_Mott_lobes_at_different_T_two_sublattice_MFT}).
Since the underlying finite temperature mean-field theory is quite
general, this approach can be straightforwardly applied to efficiently
investigate the finite temperature properties of related systems with
phases possessing nontrivial spatial structures. 

\section{Model Hamiltonian and limitations of the straightforward mean-field
decoupling approach}

\subsection{Model and its straightforward mean-field decoupled Hamiltonian }

To proceed with the discussion on a concrete basis, we consider the
simplest single-band Bose-Hubbard model \citep{Fisher_PRB_1989} on
a two-dimensional (2D) square lattice which is widely used in describing
the physics of ultracold gases in optical lattice \citep{Bloch_RMP_2008}.
Its explicit form reads
\begin{equation}
\hat{H}=-t\sum_{\langle\mathbf{i},\mathbf{j}\rangle}\left(\hat{b}_{\mathbf{i}}^{\dagger}\hat{b}_{\mathbf{j}}+\mathrm{h.c.}\right)+\sum_{\mathbf{i}}\left(\frac{U}{2}\hat{n}_{\mathbf{i}}(\hat{n}_{\mathbf{i}}-1)-\mu\hat{n}_{\mathbf{i}}\right),\label{eq:BHM}
\end{equation}
where $\hat{b}_{\mathbf{i}}^{\dagger}$ ($\hat{b}_{\mathbf{i}}$)
is the bosonic creation (annihilation) operator for the Wannier state
on the site $\mathbf{i}$ in the lowest band, $\hat{n}_{\mathbf{i}}\equiv\hat{b}_{\mathbf{i}}^{\dagger}\hat{b}_{\mathbf{i}}$
is the particle number operator. Here, $t$ is the positive hopping
amplitude of bosons between nearest neighboring lattice sites denoted
by $\langle\mathbf{i},\mathbf{j}\rangle$, $U$ is the strength of
the on-site interaction and $\mu$ is the chemical potential. 

To investigate the properties of the system, straightforward mean-field
decoupling for the hopping term is frequently employed \citep{Oosten_PRA_2001,Sachdev_QPT_2011}.
It is done by plugging the decomposition $\hat{b}_{\mathbf{i}}=\psi_{\mathbf{i}}+\left(\hat{b}_{\mathbf{i}}-\psi_{\mathbf{i}}\right)$
into the hopping term of $\hat{H}$ and keeping only up to the first-order
terms in $(\hat{b}_{\mathbf{i}}-\psi_{\mathbf{i}})$. This gives rise
to the straightforward mean-field decoupled (SMFD) Hamiltonian $\hat{H}_{\mathrm{\mathrm{SMFD}}}(\{\psi_{\mathbf{i}}^{*},\psi_{\mathbf{i}}\})$,
the explicit form of which reads
\begin{align}
 & \hat{H}_{\mathrm{\mathrm{SMFD}}}(\{\psi_{\mathbf{i}}^{*},\psi_{\mathbf{i}}\})=t\sum_{\langle\mathbf{i},\mathbf{j}\rangle}\left(\psi_{\mathbf{i}}^{*}\psi_{\mathbf{j}}+\psi_{\mathbf{i}}\psi_{\mathbf{j}}^{*}\right)\label{eq:H_SMFD_general}\\
 & -t\sum_{\langle\mathbf{i},\mathbf{j}\rangle}\left(\psi_{\mathbf{i}}^{*}\hat{b}_{\mathbf{j}}+\hat{b}_{\mathbf{i}}^{\dagger}\psi_{\mathbf{j}}+\mathrm{h.c.}\right)+\sum_{\mathbf{i}}\left[\frac{U}{2}\hat{n}_{\mathbf{i}}(\hat{n}_{\mathbf{i}}-1)-\mu\hat{n}_{\mathbf{i}}\right],\nonumber 
\end{align}
Here, $\psi_{\mathbf{i}}$ is a complex variable and assumes the physical
interpretation of the local superfluid (SF) order parameter. At zero
temperature, the value of $\psi_{\mathbf{i}}$ is expected to be determined
by minimizing the ground state energy of $\hat{H}_{\mathrm{SMFD}}(\{\psi_{\mathbf{i}}^{*},\psi_{\mathbf{i}}\})$.

\subsection{Limitations of the straightforward mean-field decoupling approach}

When utilizing the straightforward mean-field decoupled Hamiltonian
$\hat{H}_{\mathrm{\mathrm{SMFD}}}(\{\psi_{\mathbf{i}}^{*},\psi_{\mathbf{i}}\})$,
a homogeneous ansatz for $\psi_{\mathbf{i}}$, i.e., $\psi_{\mathbf{i}}=\psi$
for $\forall\mathbf{i}$ is usually further assumed (see for instance
Refs.~\citep{Oosten_PRA_2001,Sachdev_QPT_2011}). Under this homogeneous
ansatz, $\hat{H}_{\mathrm{\mathrm{SMFD}}}$ assumes the form 

\begin{align}
\hat{H}_{\mathrm{\mathrm{SMFD}}}(\psi^{*},\psi)= & 2N_{s}tz\psi^{*}\psi-\sum_{\mathbf{i}}2zt(\hat{b}_{\mathbf{i}}^{\dagger}\psi+\mathrm{h}.\mathrm{c}.)\nonumber \\
 & +\sum_{\mathbf{i}}\left[\frac{U}{2}\hat{n}_{\mathbf{i}}(\hat{n}_{\mathbf{i}}-1)-\mu\hat{n}_{\mathbf{i}}\right],\label{eq:H_SMFD_homogeneous}
\end{align}
with $z\equiv\sum_{\mathbf{j}=\langle\mathbf{i}\rangle}=4$ being
the coordination number of the 2D square lattice and $N_{s}$ being
the total number of the lattice sites. Direct calculations based on
$\hat{H}_{\mathrm{\mathrm{SMFD}}}(\psi^{*},\psi)$ can give reasonable
ground state properties of the system \citep{Oosten_PRA_2001,Sachdev_QPT_2011}
and are consistent with related results from other mean-field approaches,
for instance, bosonic Gutzwiller variational wave function approach
\citep{Krauth_PRB_1992,Jaksch_PRL_1998}. In this regard, one would
naturally expect this straightforward mean-field decoupling approach
with a generic ansatz for the mean-field $\psi_{\mathbf{i}}$ that
takes into account its possible spatial dependence, should also be
able to give reasonable predictions. 

Somewhat unexpected, one actually finds this natural expectation is
not true in general. To see this point, let us first take a simple
inhomogeneous ansatz for $\psi_{\mathbf{i}}$ as an example, where
$\psi_{\mathbf{i}}$ assumes the same value on each of the two sub-lattices
of the 2D square lattice and can assume different values on different
sub-lattice, i.e., $\psi_{\mathbf{i}}=\psi_{\sigma}$ if $\mathbf{i}\in\mathring{\sigma}$
with $\sigma=e,o$. Here, $\mathring{\sigma}$ denotes the set of
all lattice sites of the sub-lattice with the index $\sigma$, and
we denote two sub-lattice indices as $e$ and $o$. The straightforward
mean-field decoupled Hamiltonian based on this ansatz can be directly
obtained. Its explicit form reads 
\begin{align}
 & \hat{H}_{\mathrm{SMFD}}(\{\psi_{\sigma}^{*},\psi_{\sigma}\})=N_{s}zt(\psi_{e}^{*}\psi_{o}+\psi_{o}^{*}\psi_{e})\label{eq:H_SMFD_two_sub_lat}\\
 & +\sum_{\sigma}\sum_{\mathbf{i}\in\mathring{\sigma}}\left\{ \frac{1}{2}U_{s}\hat{n}_{\mathbf{i}}(\hat{n}_{\mathbf{i}}-1)-\mu\hat{n}_{\mathbf{i}}-2zt\left(\hat{b}_{\mathbf{i}}^{\dagger}\psi_{\bar{\sigma}}+\mathrm{h}.\mathrm{c}.\right)\right\} ,\nonumber 
\end{align}
where $\bar{\sigma}$ is defined by $\bar{\sigma}=o$ if $\sigma=e$
and $\bar{\sigma}=e$ if $\sigma=o$. We notice that the first term
of $\hat{H}_{\mathrm{SMFD}}(\{\psi_{\sigma}^{*},\psi_{\sigma}\})$
is a quadrature with respect to $\{\psi_{\sigma}^{*},\psi_{\sigma}\}$
that is not positive definite, i.e., $N_{s}zt(\psi_{e}^{*}\psi_{o}+\psi_{o}^{*}\psi_{e})=(\psi_{e}^{*},\psi_{o}^{*})\left(\begin{array}{cc}
0 & N_{s}zt\\
N_{s}zt & 0
\end{array}\right)(\psi_{e},\psi_{o})^{T}$ with $\left(\begin{array}{cc}
0 & N_{s}zt\\
N_{s}zt & 0
\end{array}\right)$ clearly not being a positive definite matrix. This directly indicates
that $\hat{H}_{\mathrm{SMFD}}(\{\psi_{\sigma}^{*},\psi_{\sigma}\})$
generally assumes no lower energy bound with respect to $\{\psi_{\sigma}^{*},\psi_{\sigma}\}$,
hence making the mean-field theory Eq.~(\ref{eq:H_SMFD_two_sub_lat})
not reliable anymore. 

More generally, we notice from Eq.~(\ref{eq:H_SMFD_general}) that
the straightforward mean-field decoupled Hamiltonian $\hat{H}_{\mathrm{\mathrm{SMFD}}}(\{\psi_{\mathbf{i}}^{*},\psi_{\mathbf{i}}\})$
is not guaranteed to assume a lower energy bound with respect to $\{\psi_{\mathbf{i}}^{*},\psi_{\mathbf{i}}\}$,
since its first term is a quadrature of $\{\psi_{\mathbf{i}}^{*},\psi_{\mathbf{i}}\}$
that is not guaranteed to be positive definite. Although this may
not cause serious issues in investigating zero temperature properties
in related systems \citep{Sheshadri_PRL_1995,Kurdestany_Ann_Phys_Berlin_2012,Pai_PRB_2012,Puschmann_Ann_Phys_2021}
since iterative self-consistent approaches are usually employed, the
lack of the lower energy bound is fatal in the finite temperature
case due to the presence of the thermal fluctuations of the order
parameter field. This thus indicates the reliability of the mean-field
theory Eq.~(\ref{eq:H_SMFD_general}) is limited to the applications
combined with the ansatzes for $\psi_{\mathbf{i}}$ that give positive
definite quadratures of $\{\psi_{\mathbf{i}}^{*},\psi_{\mathbf{i}}\}$.
In this regard, it is desirable that a reliable construction for the
mean-field theory beyond this limitation can be established. Indeed,
as we shall present in the following, such a finite temperature mean-field
theory that can reliably work together with generic ansatzes for $\psi_{\mathbf{i}}$
can be established via formulating the exact partition function of
the system in the functional integral form and taking the classical
limit of the quantum order parameter field introduced by the standard
Hubbard-Stratonovich transformation \citep{Hubbard_PRL_1959,Stratonovich_HST_1957,Altland_CFT_2010}.

\section{Finite temperature mean-field theory with intrinsic non-hermitian
structure }

To solve the issue of the mean-field theory constructed by the straightforward
mean-field decoupling approach, we first write down the exact partition
function of the system $Z=\mathrm{tr}[e^{-\beta\hat{H}}]$ in the
standard coherent state functional integral formulation, the explicit
form of which reads 
\begin{equation}
Z=\int\mathcal{D}(\{b_{\mathbf{i}}^{*}(\tau),b_{\mathbf{i}}(\tau)\})\,e^{-S[\{b_{\mathbf{i}}^{*}(\tau),b_{\mathbf{i}}(\tau)\}]},
\end{equation}
with the action $S[\{b_{\mathbf{i}}^{*}(\tau),b_{\mathbf{i}}(\tau)\}]$
assuming the explicit form
\begin{align}
 & S[\{b_{\mathbf{i}}^{*}(\tau),b_{\mathbf{i}}(\tau)\}]=\int_{0}^{\beta}d\tau\left\{ -t\sum_{\langle\mathbf{i},\mathbf{j}\rangle}\left(b_{\mathbf{i}}^{*}b_{\mathbf{j}}+\mathrm{c.c.}\right)\right.\nonumber \\
 & \left.+\sum_{\mathbf{i}}\left(b_{\mathbf{i}}^{*}\partial_{\tau}b_{\mathbf{i}}+\frac{U}{2}b_{\mathbf{i}}^{*}b_{\mathbf{i}}^{*}b_{\mathbf{i}}b_{\mathbf{i}}-\mu b_{\mathbf{i}}^{*}b_{\mathbf{i}}\right)\right\} ,\label{eq:Exact_action}
\end{align}
where $b_{\mathbf{i}}^{*}(\tau)$ ($b_{\mathbf{i}}(\tau)$) is the
complex field that corresponds to the bosonic operator $\hat{b}_{\mathrm{i}}^{\dagger}$
($\hat{b}_{\mathrm{i}}$). Via the standard Hubbard-Stratonovich transformation
(HST) \citep{Hubbard_PRL_1959,Stratonovich_HST_1957,Altland_CFT_2010},
the fluctuating quantum superfluid order parameter field $\psi_{\mathbf{i}}(\tau)$
can be directly introduced into the partition function to decouple
the hopping term in the action as what has been done routinely in
literature (see, for instance, Refs.~\citep{Fisher_PRB_1989,Sachdev_QPT_2011}).
However, we remark here that one should pay special attention to the
convergence of the complex Gaussian integral involved in the HST,
which is frequently overlooked in literature, since the ``hopping
matrix'' (to be defined explicitly below) is generically an indefinite
matrix.

\subsection{Hubbard-Stratonovich transformation in the presence of the indefinite
hopping matrix}

To perform the HST in the presence of the indefinite hopping matrix,
let us first reformulate the hopping term in the action, i.e., $-t\sum_{\langle\mathbf{i},\mathbf{j}\rangle}\left(b_{\mathbf{i}}^{*}b_{\mathbf{j}}+\mathrm{c.c.}\right)$,
into a matrix form, i.e., 
\begin{equation}
-t\sum_{\langle\mathbf{i},\mathbf{j}\rangle}\left(b_{\mathbf{i}}^{*}b_{\mathbf{j}}+\mathrm{c.c.}\right)=\sum_{\mathbf{i},\mathbf{j}}b_{\mathbf{i}}^{*}\mathbf{T}_{\mathbf{i}\mathbf{j}}b_{\mathbf{j}}=B^{\dagger}\mathbf{T}B,\label{eq:hopping_term_before_HST}
\end{equation}
where $\mathbf{T}$ is the ``hopping matrix'' with its matrix elements
denoted by $\mathbf{T}_{\mathbf{i}\mathbf{j}}$, $B$ is an $N_{s}$
dimensional column vector that collects all $b_{\mathbf{i}}$, i.e.,
$B\equiv(\cdots,b_{\mathbf{i}},\cdots)^{T}$. Since $\mathbf{T}$
is indefinite in general, the quadrature in Eq.~(\ref{eq:hopping_term_before_HST})
is indefinite too. We separate the positive definite part of the quadrature
from its negative definite part by first diagonalizing $\mathbf{T}$
with a unitary matrix $U$, i.e., $U\mathbf{T}U^{\dagger}=E_{(+)}\bigoplus E_{(-)}$,
where $E_{(+)}$ ($E_{(-)}$) is an $N^{+}$ ($N^{-}$) dimensional
diagonal matrix that contains all the $N_{+}$ ($N_{-}$) positive
(negative) eigenvalues of $\mathbf{T}$. Then, the hopping term $B^{\dagger}\mathbf{T}B=\tilde{B}^{\dagger}(E_{(+)}\bigoplus E_{(-)})\tilde{B}$
with $\tilde{B}\equiv UB$ and can be further written as a sum of
the positive and the negative definite part, i.e., 

\begin{align}
B^{\dagger}\mathbf{T}B & =\tilde{B}_{(+)}^{\dagger}E_{(+)}\tilde{B}_{(+)}+\tilde{B}_{(-)}^{\dagger}E_{(-)}\tilde{B}_{(-)}\label{eq:qudrature_of_hopping_positive_negative_separation}
\end{align}
where $\tilde{B}_{(+)}$ ($\tilde{B}_{(-)}$) is an $N_{+}$ ($N_{-}$)
dimensional column vector that contains the first $N_{+}$ (last $N_{-}$)
elements of $\tilde{B}$, i.e., $\tilde{B}_{(+)}=\left(\begin{array}{cc}
\mathbf{I}_{(+)} & \mathbf{0}_{(+)}\end{array}\right)\tilde{B}$ and $\tilde{B}_{(-)}=\left(\begin{array}{cc}
\mathbf{0}_{(-)} & \mathbf{I}_{(-)}\end{array}\right)\tilde{B}$, with $\mathbf{I}_{(\pm)}$ being an $N_{\pm}$ dimensional identity
matrix and $\mathbf{0}_{(\pm)}$ being an $N_{\pm}\times N_{\mp}$
zero matrix. 

The HST of the whole hopping term is facilitated by performing two
standard HST for the positive definite part and the negative definite
part, separately. For the positive definite part, the HST reads
\begin{align}
 & e^{-\tilde{B}_{(+)}^{\dagger}E_{(+)}\tilde{B}_{(+)}}\label{eq:HST_positive_part}\\
= & \int d(\tilde{\Psi}_{(+)}^{\dagger},\tilde{\Psi}_{(+)})e^{-\tilde{\Psi}_{(+)}^{\dagger}E_{(+)}\tilde{\Psi}_{(+)}+i\left(\tilde{\Psi}_{(+)}^{\dagger}E_{(+)}\tilde{B}_{(+)}+\mathrm{c.c.}\right)},\nonumber 
\end{align}
while for the negative definite part, the corresponding HST reads
\begin{align}
 & e^{-\tilde{B}_{(-)}^{\dagger}E_{(-)}\tilde{B}_{(-)}}\label{eq:HST_negative_part}\\
 & =\int d(\tilde{\Psi}_{(-)}^{\dagger},\tilde{\Psi}_{(-)})e^{\tilde{\Psi}_{(-)}^{\dagger}E_{(-)}\tilde{\Psi}_{(-)}+\left(\tilde{\Psi}_{(-)}^{\dagger}E_{(-)}\tilde{B}_{(-)}+\mathrm{c.c.}\right)},\nonumber 
\end{align}
with $\tilde{\Psi}_{(\pm)}$ being an $N_{\pm}$ dimensional complex
vector variable and $d(\tilde{\Psi}_{(\pm)}^{\dagger},\tilde{\Psi}_{(\pm)})\equiv\pi^{-N_{\pm}}\prod_{i=1}^{N_{\pm}}E_{(\pm)i}d(\tilde{\psi}_{(\pm)i}^{*},\tilde{\psi}_{(\pm)i})$.
In particular, we notice from Eq.~(\ref{eq:HST_positive_part}) that
the imaginary unit $i$ appears as an overall prefactor of the coupling
term between $\tilde{\Psi}_{(+)}$ and $\tilde{B}_{(+)}$ in the HST
for the positive definite part, while this is not the case for the
negative definite one. 

With the two separate HST in Eqs.~(\ref{eq:HST_positive_part},~\ref{eq:HST_negative_part}),
we can straightforwardly express the hopping term in the partition
function as an integral with respect to the $N_{s}$ dimensional complex
vector variable $\Psi\equiv U^{\dagger}(\tilde{\Psi}_{(+)}^{T}\tilde{\Psi}_{(-)}^{T})^{T}$,
i.e., 
\begin{equation}
e^{-B^{\dagger}\mathbf{T}B}=\int d(\Psi^{\dagger},\Psi)e^{-\Psi^{\dagger}\mathbf{T}_{\mathrm{H}}\Psi+\left(\Psi^{\dagger}\mathbf{T}_{\mathrm{NH}}B+B^{\dagger}\mathbf{T}_{\mathrm{NH}}\Psi\right)},\label{eq:hopping_term_complete_HST}
\end{equation}
where 
\begin{align}
\mathbf{T}_{\mathrm{H}} & \equiv U^{\dagger}\left(E_{(+)}\bigoplus(-E_{(-)})\right)U,\label{eq:T_H}\\
\mathbf{T}_{\mathrm{NH}} & \equiv U^{\dagger}\left((iE_{(+)})\bigoplus(E_{(-)})\right)U.\label{eq:T_NH}
\end{align}
In particular, we notice that $T_{\mathrm{H}}$ is a positive definite
hermitian matrix, while $\mathbf{T}_{\mathrm{NH}}$ is a non-hermitian
matrix in general, and gives rise to the non-hermitian structure of
the mean-field Hamiltonian as we shall see in the following.

\subsection{Non-hermitian structure of the mean-field Hamiltonian}

With the HST in Eq.~(\ref{eq:hopping_term_complete_HST}) for the
hopping term, we can straightforwardly reformulate the partition function
$Z$ as a functional integral with respect to both $B(\tau)$ and
$\Psi(\tau)$, the explicit form of which reads 
\begin{align}
Z= & \int\mathcal{D}(\Psi^{\dagger}(\tau),\Psi(\tau),B^{\dagger}(\tau),B(\tau))\label{eq:Z_after_HST}\\
 & \times e^{-S[\Psi^{\dagger}(\tau),\Psi(\tau),B^{\dagger}(\tau),B(\tau)]},\nonumber 
\end{align}
where 
\begin{align}
 & S[\Psi^{\dagger}(\tau),\Psi(\tau),B^{\dagger}(\tau),B(\tau)]=\int_{0}^{\beta}d\tau\Psi^{\dagger}T_{\mathrm{H}}\Psi\nonumber \\
 & +\int_{0}^{\beta}d\tau\sum_{\mathbf{i}}\left\{ b_{\mathbf{i}}^{*}\partial_{\tau}b_{\mathbf{i}}+\frac{U}{2}b_{\mathbf{i}}^{*}b_{\mathbf{i}}^{*}b_{\mathbf{i}}b_{\mathbf{i}}-\mu b_{\mathbf{i}}^{*}b_{\mathbf{i}}\right.\nonumber \\
 & \left.-\left[\left(\Psi^{\dagger}\mathbf{T}_{\mathrm{NH}}\right)_{\mathbf{i}}b_{\mathbf{i}}+b_{\mathbf{i}}^{*}\left(\mathbf{T}_{\mathrm{NH}}\Psi\right)_{\mathbf{i}}\right]\right\} .\label{eq:complete_action_aftere_HST}
\end{align}
Up to now, Eqs.~(\ref{eq:Z_after_HST},~\ref{eq:complete_action_aftere_HST})
essentially correspond to an exact reformulation of the quantum partition
function of the system, based on which one can formulate different
perturbative approaches following the similar way as the ones presented
in, for instance, Refs.~\citep{Fisher_PRB_1989,Sachdev_QPT_2011,Sengupta_PRA_2005}.
Here, to investigate the finite temperature properties of the system,
we shall formulate a different approach based on analyzing the minimum
of the effective potential function {[}cf.~Eq.~(\ref{eq:Omega_general}){]}
of the system constructed from its proper mean-field Hamiltonian {[}cf.~Eq.(\ref{eq:HMF_general_form}){]}.
More specifically, to proceed further, we make the mean-field (classical)
approximation that neglect the quantum fluctuations of the order parameter
field $\Psi(\tau)$, i.e., assume the superfluid order parameter field
$\Psi(\tau)$ does not depend on $\tau$, i.e., $\Psi(\tau)=\Psi$.
This gives rise to the mean-field partition function $Z_{\mathrm{MF}}$
with the explicit form
\begin{equation}
Z_{\mathrm{MF}}=\int d(\Psi^{\dagger},\Psi)\mathcal{D}(B^{\dagger}(\tau),B(\tau))e^{-S_{\mathrm{MF}}[\Psi^{\dagger},\Psi,B^{\dagger}(\tau),B(\tau)]},\label{eq:mean-field_partition_function}
\end{equation}
where the mean-field action $S_{\mathrm{MF}}$ reads
\begin{align}
 & S_{\mathrm{MF}}[\Psi^{\dagger},\Psi,B^{\dagger}(\tau),B(\tau)]=\beta\Psi^{\dagger}T_{\mathrm{H}}\Psi\nonumber \\
 & +\int_{0}^{\beta}d\tau\sum_{\mathbf{i}}\left\{ b_{\mathbf{i}}^{*}\partial_{\tau}b_{\mathbf{i}}+\frac{U}{2}b_{\mathbf{i}}^{*}b_{\mathbf{i}}^{*}b_{\mathbf{i}}b_{\mathbf{i}}-\mu b_{\mathbf{i}}^{*}b_{\mathbf{i}}\right.\nonumber \\
 & \left.-\left[\left(\Psi^{\dagger}\mathbf{T}_{\mathrm{NH}}\right)_{\mathbf{i}}b_{\mathbf{i}}+b_{\mathbf{i}}^{*}\left(\mathbf{T}_{\mathrm{NH}}\Psi\right)_{\mathbf{i}}\right]\right\} .\label{eq:mean_field_complete_action_aftere_HST}
\end{align}
Noticing that the $\tau$ dependence of the second term of the mean-field
action $S_{\mathrm{MF}}$ only comes from $\{b_{\mathbf{i}}^{*}(\tau),b_{\mathbf{i}}(\tau)\}$,
we can reformulate $Z_{\mathrm{MF}}$ as (see Appendix~\ref{sec:Transformation-from-functional-back-to-operator}
for more details) 
\begin{equation}
Z_{\mathrm{MF}}=\int d(\Psi^{\dagger},\Psi)\mathrm{tr}\left[e^{-\beta\hat{H}_{\mathrm{MF}}(\Psi^{\dagger},\Psi)}\right],\label{eq:Z_MF}
\end{equation}
where the mean-field Hamiltonian $\hat{H}_{\mathrm{MF}}(\Psi^{\dagger},\Psi)$
assumes the explicit form
\begin{equation}
\hat{H}_{\mathrm{MF}}(\Psi^{\dagger},\Psi)=\Psi^{\dagger}\mathbf{T}_{\mathrm{H}}\Psi+\sum_{\mathbf{i}}\hat{H}_{\mathrm{SS}}^{(\mathbf{i})}(\Psi^{\dagger},\Psi),\label{eq:HMF_general_form}
\end{equation}
with $\hat{H}_{\mathrm{SS}}^{(\mathbf{i})}(\Psi^{\dagger},\Psi)$
being a single site Hamiltonian that only involves operators on site
$\mathbf{i}$, the explicit form of which reads
\begin{align}
 & \hat{H}_{\mathrm{SS}}^{(\mathbf{i})}(\Psi^{\dagger},\Psi)\equiv\frac{U}{2}\hat{n}_{\mathbf{i}}(\hat{n}_{\mathbf{i}}-1)-\mu\hat{n}_{\mathbf{i}}\nonumber \\
 & -\left[\left(\Psi^{\dagger}\mathbf{T}_{\mathrm{NH}}\right)_{\mathbf{i}}\hat{b}_{\mathbf{i}}+\hat{b}_{\mathbf{i}}^{\dagger}\left(\mathbf{T}_{\mathrm{NH}}\Psi\right)_{\mathbf{i}}\right].\label{eq:HSS_general_form}
\end{align}

We notice $\hat{H}_{\mathrm{MF}}(\Psi^{\dagger},\Psi)$ is generally
a non-hermitian Hamiltonian due to the general non-hermiticity of
$\mathbf{T}_{\mathrm{NH}}$. Such a non-hermitian structure is reminiscent
of non-hermitian Hamiltonians widely used to describe physics of open
quantum systems (see Ref.~\citep{Ashida_Adv_Phys_2020} for a recent
review), where the non-hermiticity originates from dissipations. In
contrast, here, it originates from the hermitian hopping term with
an indefinite hopping matrix. It is worth mentioning that although
the non-hermitian structure of the mean-field Hamiltonian manifests
itself explicitly in concrete calculations via giving rise to complex
eigenvalues, this non-hermiticity does not give rise to any physical
effect characteristic of open quantum many-body systems (cf.~Sec.~\ref{subsec:Two-sublattice-FTMFT}
for a detailed and concrete discussion). Moreover, we would like to
remark that the above derivation of the mean-field Hamiltonian is
completely general and can be directly applied to systems with complex
hopping terms, for instance, systems with spin-orbit coupling \citep{Cole_PRL_2012,Radic_PRL_2012,Cai_PRA_2012,Hickey_PRL_2014,He_PRA_2015,Wu_Science_2016},
effective magnetic flux \citep{Cooper_PRL_2011,Aidelsburger_PRL_2011,Duca_Science_2015},
etc.

\subsection{Finite temperature mean-field theory in terms of the potential function
for the order parameter field}

To make the mean-field theory Eq.~(\ref{eq:Z_MF}) a useful and efficient
tool for investigating both zero temperature and finite temperature
properties of the system, we further reformulate the integrand in
the expression for $Z_{\mathrm{MF}}$ in Eq.~(\ref{eq:Z_MF}) as
an exponential of a potential function of the superfluid order parameter
field alone, i.e., 
\begin{equation}
Z_{\mathrm{MF}}=\int d(\Psi^{\dagger},\Psi)e^{-\beta N_{s}\Omega_{t,U,\mu,\beta}(\Psi^{\dagger},\Psi)},\label{eq:Z_MF_Omega}
\end{equation}
where the potential function $\Omega_{t,U,\mu,\beta}(\Psi^{\dagger},\Psi)$
reads
\begin{align}
 & \Omega_{t,U,\mu,\beta}(\Psi^{\dagger},\Psi)\nonumber \\
= & \frac{1}{N_{s}}\left(\Psi^{\dagger}\mathbf{T}_{\mathrm{H}}\Psi-\frac{1}{\beta}\ln\prod_{\mathbf{i}}\mathrm{tr}\left[e^{-\beta\hat{H}_{\mathrm{SS}}^{(\mathbf{i})}(\Psi^{\dagger},\Psi)}\right]\right).\label{eq:Omega_general}
\end{align}

The most appealing feature of this reformulation is that we can analyze
the properties of the system by simply investigating the saddle points
of the potential function $\Omega_{t,U,\mu,\beta}(\Psi^{\dagger},\Psi)$.
This is due to the fact that in the thermodynamic limit ($N_{s}\rightarrow+\infty$),
the partition function $Z_{\mathrm{MF}}$ is exactly determined by
the saddle point value of $\Omega_{t,U,\mu,\beta}(\Psi^{\dagger},\Psi)$,
hence the value of the SF order parameter $\bar{\Psi}$ at temperature
$T$ is determined by the value of $\Psi$ that minimize $\Omega_{t,U,\mu,\beta}(\Psi^{\dagger},\Psi)$.
Although the explicit form of $\Omega_{t,U,\mu,\beta}(\Psi^{\dagger},\Psi)$
can not be obtained analytically due to the trace in its expression,
its value at given $(\Psi^{\dagger},\Psi)$ can be calculated very
efficiently at sufficiently high accuracy by employing a large enough
cut-off $n_{\mathrm{max}}$ in the dimension of the local Hilbert
space for the site $\mathbf{i}$. For the temperature regime of most
interests, where thermal fluctuations compete strongly with the two
other energy scales in the system (the ones associated with the on-site
interaction and the hopping), i.e., $k_{B}T\sim U\sim zt$, a cut-off
$n_{\mathrm{max}}$ of $\mathcal{O}(10)$ is already large enough. 

\subsection{Finite temperature mean-field theory combined with the homogeneous
and the inhomogeneous two-sublattice ansatz for $\psi_{\mathbf{i}}$}

To illustrate concretely how the finite temperature mean-field theory
is applied, we used it together with the homogeneous and an inhomogeneous
two-sublattice ansatz to investigate the superfluid transition of
the system at finite temperature. To this end, let us first diagonalize
the hopping term that appears in the action in Eq.~(\ref{eq:Exact_action})
by a straightforward lattice Fourier transformation, i.e., 
\begin{align}
-t\sum_{\langle\mathbf{i},\mathbf{j}\rangle}\left(b_{\mathbf{i}}^{*}b_{\mathbf{j}}+\mathrm{c}.\mathrm{c}.\right) & =\sum_{\mathbf{k}}[\mathcal{E}_{(+)}(\mathbf{k})+\mathcal{E}_{(-)}(\mathbf{k})]\tilde{b}_{\mathbf{k}}^{*}\tilde{b}_{\mathbf{k}},\label{eq:hopping_term_diagonal_form}
\end{align}
where $\tilde{b}_{\mathbf{k}}\equiv\left(L_{x}L_{y}\right)^{-1/2}\sum_{\mathbf{i}}e^{-i\mathbf{k}\cdot\mathbf{i}}b_{\mathbf{i}},$
$\mathcal{E}_{(\pm)}(\mathbf{k})\equiv\theta(\pm\mathcal{E}(\mathbf{k}))\mathcal{E}(\mathbf{k})$,
$\mathcal{E}(\mathbf{k})\equiv-4t\left(\cos k_{x}+\cos k_{y}\right)$,
$\theta(x)$ is the Heaviside step function, i.e., $\theta(x)=1$,
if $x>0$, and $\theta(x)=0$, if $x<0$, and $k_{\alpha}=0\cdot2\pi/L_{\alpha},1\cdot2\pi/L_{\alpha},2\cdot2\pi/L_{\alpha},\ldots,(L_{\alpha}-1)\cdot2\pi/L_{\alpha}$,
with $\alpha=x,y$ and $L_{\alpha}$ being the number of lattice sites
along the direction $\alpha$. This enables us to write down the explicit
form of the mean-field hamiltonian $\hat{H}_{\mathrm{MF}}(\Psi^{\dagger},\Psi)$
(see Appendix \ref{App:Explicit_form_HMF} for derivation details),
which reads
\begin{align}
 & \hat{H}_{\mathrm{MF}}(\Psi^{\dagger},\Psi)=\sum_{\mathbf{i}}\left(\frac{U}{2}\hat{n}_{\mathbf{i}}(\hat{n}_{\mathbf{i}}-1)-\mu\hat{n}_{\mathbf{i}}\right)\label{eq:HMF_tight_binding_general_form}\\
 & +\frac{1}{N_{s}}\sum_{\mathbf{i},\mathbf{i}',\mathbf{k}}\psi_{\mathbf{i}}^{*}\left([\theta(\mathcal{E}(\mathbf{k}))-\theta(-\mathcal{E}(\mathbf{k}))]\mathcal{E}(\mathbf{k})e^{i\mathbf{k}\cdot(\mathbf{i}-\mathbf{i}')}\right)\psi_{\mathbf{i}'}\nonumber \\
 & +\frac{1}{N_{s}}\sum_{\mathbf{i},\mathbf{i}',\mathbf{k}}\psi_{\mathbf{i}}^{*}\left([\theta(-\mathcal{E}(\mathbf{k}))+i\theta(\mathcal{E}(\mathbf{k}))]\mathcal{E}(\mathbf{k})e^{i\mathbf{k}\cdot(\mathbf{i}-\mathbf{i}')}\right)\hat{b}_{\mathbf{i}'}\nonumber \\
 & +\frac{1}{N_{s}}\sum_{\mathbf{i},\mathbf{i}',\mathbf{k}}\hat{b}_{\mathbf{i}}^{\dagger}\left([\theta(-\mathcal{E}(\mathbf{k}))+i\theta(\mathcal{E}(\mathbf{k}))]\mathcal{E}(\mathbf{k})e^{i\mathbf{k}\cdot(\mathbf{i}-\mathbf{i}')}\right)\psi_{\mathbf{i}'}.\nonumber 
\end{align}

\subsubsection{Homogeneous finite temperature mean-field theory}

Now we can easily construct the simplest mean field theory by assuming
that the superfluid order parameter field is homogenous, i.e., $\psi_{\mathbf{i}}=\psi$.
Plugging this homogeneous ansatz for $\psi_{\mathbf{i}}$ into Eq.~(\ref{eq:HMF_tight_binding_general_form}),
one can directly obtain (see Appendix \ref{App:Explicit_form_HMF}
for derivation details)
\begin{align}
 & \hat{H}_{\mathrm{MF}}(\psi^{*},\psi)=\sum_{\mathbf{i}}\left[2tz\psi^{*}\psi+\hat{H}_{\mathrm{SS}}^{(\mathbf{i})}(\psi^{*},\psi)\right],\label{eq:HMF_homogeneous_ansatz}
\end{align}
with 
\begin{equation}
\hat{H}_{\mathrm{SS}}^{(\mathbf{i})}(\psi^{*},\psi)=-2zt(\hat{b}_{\mathbf{i}}^{\dagger}\psi+\mathrm{h}.\mathrm{c}.)+\frac{U}{2}\hat{n}_{\mathbf{i}}(\hat{n}_{\mathbf{i}}-1)-\mu\hat{n}_{\mathbf{i}},
\end{equation}
The corresponding mean-field partition function $Z_{\mathrm{MF}}$
in terms of the potential function $\Omega_{t,U,\mu,\beta}(\psi^{*},\psi)$
reads 
\begin{align}
Z_{\mathrm{MF}} & =\int d(\psi^{*},\psi)e^{-\beta N_{s}\Omega_{t,U,\mu,\beta}(\psi^{*},\psi)},\label{eq:ZMF_homogeneous_ansatz}
\end{align}
with
\begin{equation}
\Omega_{t,U,\mu,\beta}(\psi^{*},\psi)=2tz\psi^{*}\psi-\frac{1}{\beta}\ln\mathrm{tr}\left[e^{-\beta\hat{H}_{\mathrm{SS}}(\psi^{*},\psi)}\right].\label{eq:Omega_homogeneous_ansatz}
\end{equation}
The trace in Eq.~(\ref{eq:Omega_homogeneous_ansatz}) can be calculated
by first diagonalizing the single-site Hamiltonian $\hat{H}_{\mathrm{SS}}(\psi^{*},\psi)$
in the local occupation number basis with a cut-off $n_{\mathrm{max}}$
for the occupation number, and then performing the summation $\sum_{n=0}^{n_{\max}}e^{-\beta\varepsilon_{n}(\psi^{*},\psi)}$,
with $\varepsilon_{n}(\psi^{*},\psi)$ being the eigenenergy of $\hat{H}_{\mathrm{SS}}(\psi^{*},\psi)$.
This enables us to extract the dependence of $\Omega_{t,U,\mu,\beta}(\psi^{*},\psi)$
on $\psi$ at generic fixed system parameters $\{t,U,\mu,\beta\}$,
and to obtain the saddle point. 

\begin{figure}
\includegraphics[width=1.7in]{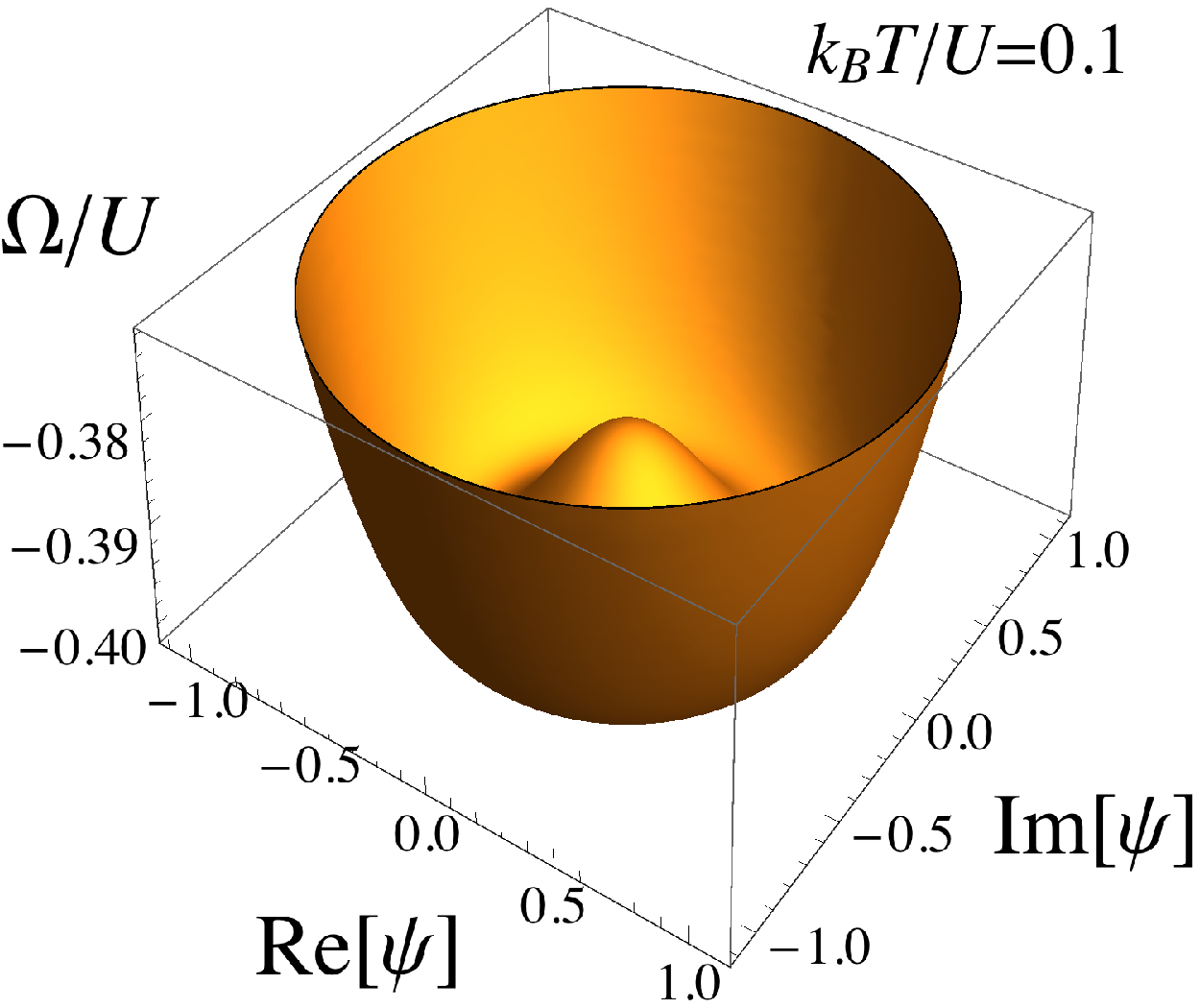}\includegraphics[width=1.7in]{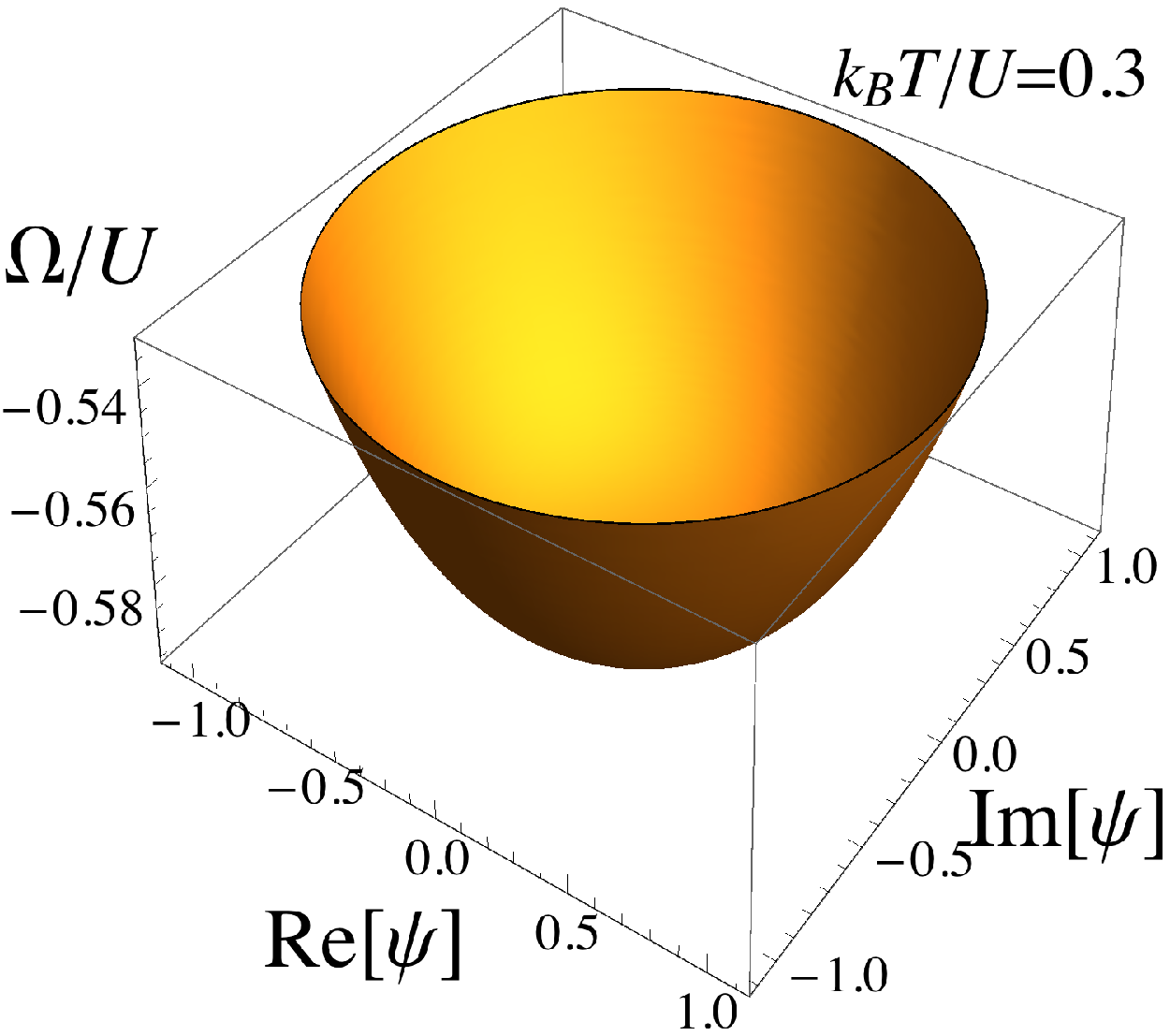}

\includegraphics[width=1.6in]{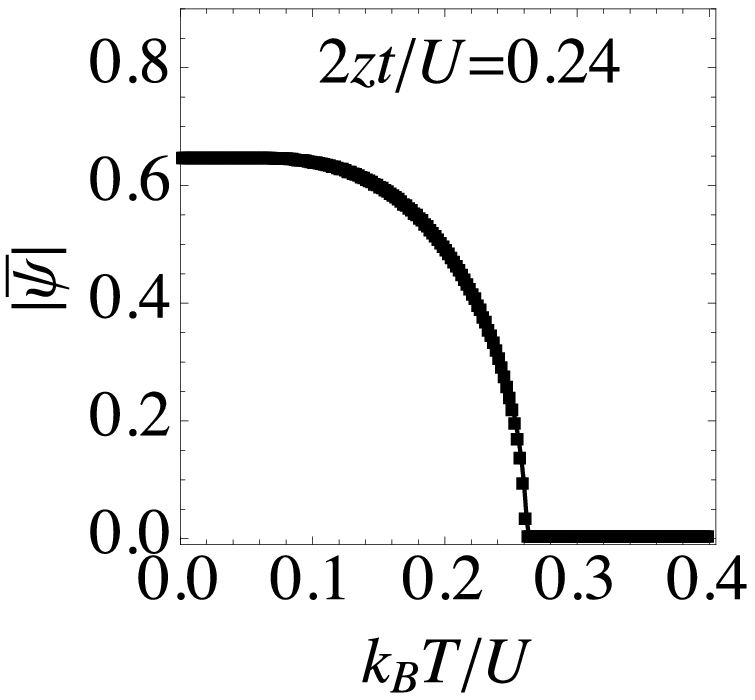}\includegraphics[width=1.8in]{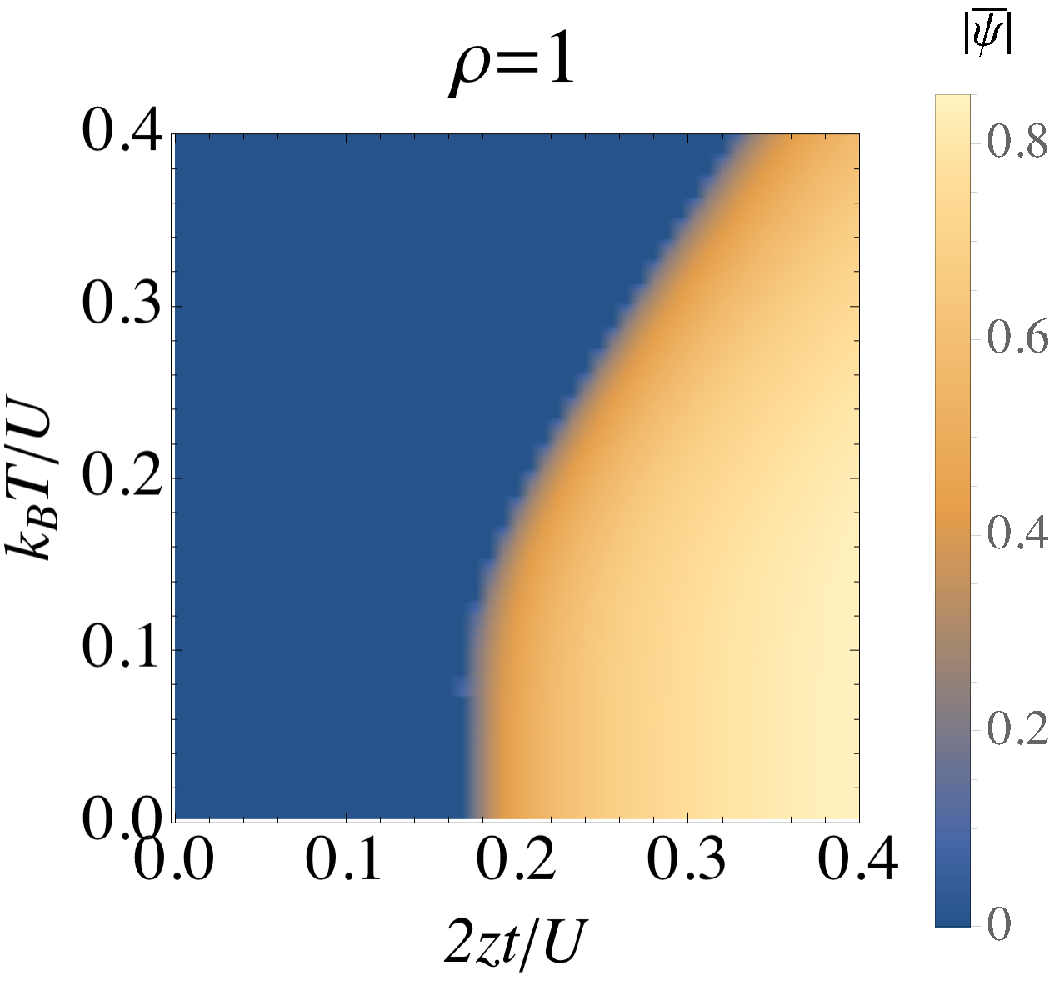}

\caption{Upper panels: Two typical landscapes of the potential function $\Omega_{t,U,\mu,\beta}(\psi^{*},\psi)$
in the homogeneous finite temperature mean-field theory at different
temperatures. Other system parameters are kept the same in these two
plots, with $2zt/U=0.24$ and the filling factor $\rho\equiv N_{s}^{-1}\sum_{\mathbf{i}}\langle\hat{n}_{\mathbf{i}}\rangle=1$.
Lower-left panel: Superfluid order parameter $|\bar{\psi}|$ as a
function of the temperature at unit filling ($\rho=1$) with $2zt/U=0.24$.
Lower-right panel: Finite temperature phase diagram of the system
at unit filling ($\rho=1$). See text for more details.}
\label{Fig_1_Finite_T_Phase_diagram}
\end{figure}

In the upper panels of Fig.~\ref{Fig_1_Finite_T_Phase_diagram},
we show two typical landscapes of the potential function $\Omega_{t,U,\mu,\beta}(\psi^{*},\psi)$
at different temperatures (other system parameters are kept the same
with $2zt/U=0.24$ and the filling factor $\rho\equiv N_{s}^{-1}\sum_{\mathbf{i}}\langle\hat{n}_{\mathbf{i}}\rangle=1$).
We see that at the low temperature ($k_{B}T/U=0.1$ in this case)
the potential function assumes a typical Mexican hat structure with
its minimums located on a ring where $\psi$ assumes non-zero modulus,
while at the high temperature ($k_{B}T/U=0.3$ in this case) the potential
function assumes a bowl structure with its unique minimum located
at $\psi=0$. This kind of change in the structure of the potential
function directly indicates the temperature-driven phase transition
from the superfluid to the normal phase is a second-order transition.
Indeed, from the lower-left panel of Fig.~\ref{Fig_1_Finite_T_Phase_diagram},
we see that the modulus of the superfluid order parameter $|\bar{\psi}|$
of the system continuously decreases to zero as $T$ increases. The
complete finite temperature phase diagram of the system at unit filling
($\rho=1$) is shown in the lower-right panel of Fig.~\ref{Fig_1_Finite_T_Phase_diagram},
where we notice that the critical temperature $T_{C}$ for the superfluid
to normal phase transition only show obvious increasing behavior with
respect to the hopping amplitude once the two competing energy scales,
i.e., $2zt$ and $k_{B}T$, are comparable with each other. 

\begin{figure}
\includegraphics[width=1.7in]{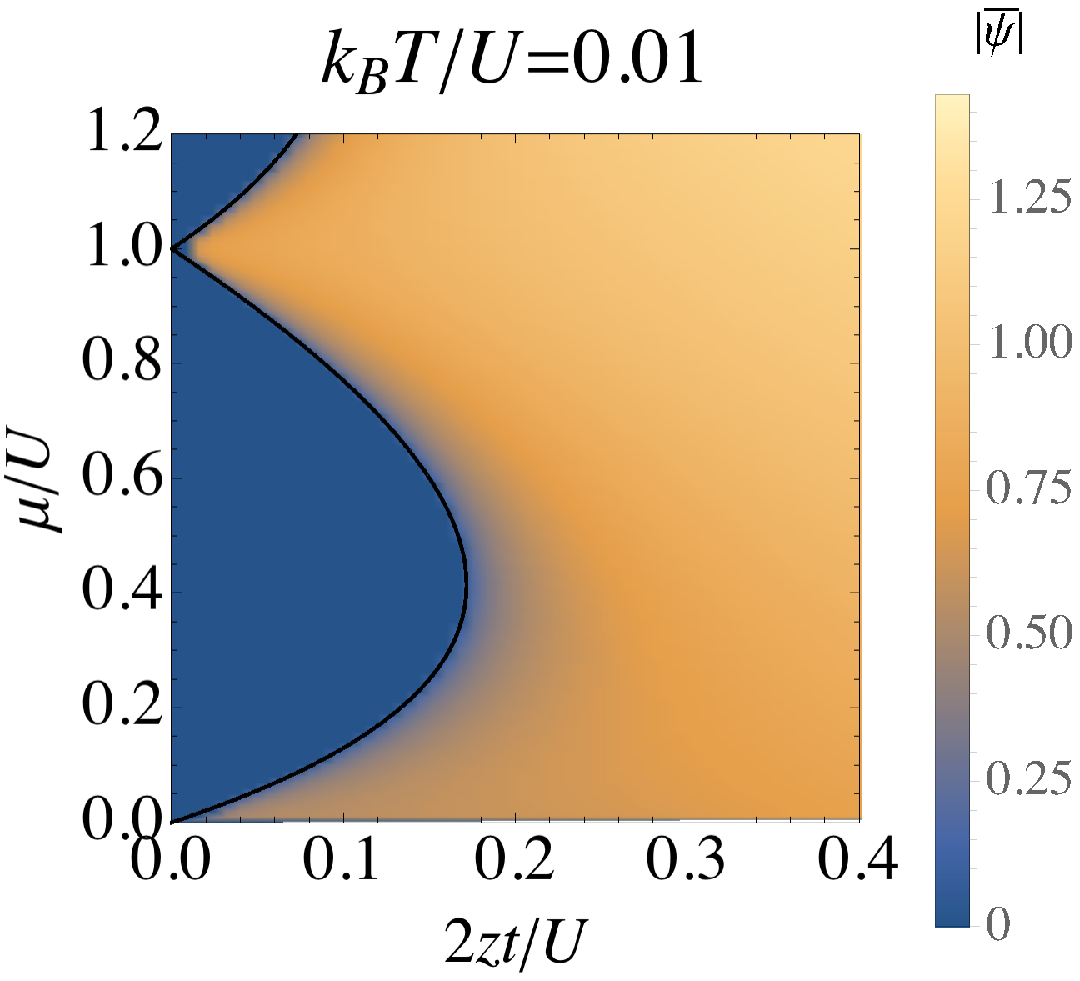}\includegraphics[width=1.7in]{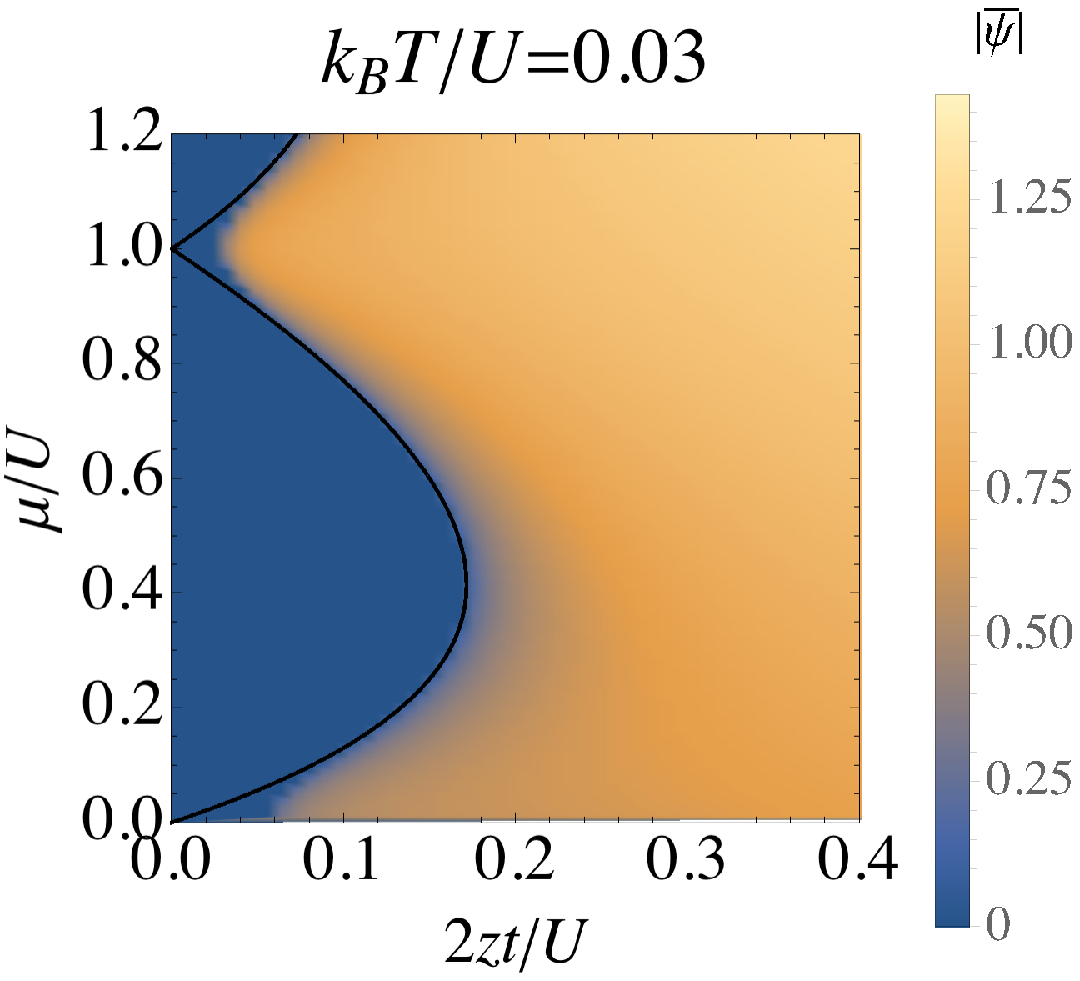}

\includegraphics[width=1.7in]{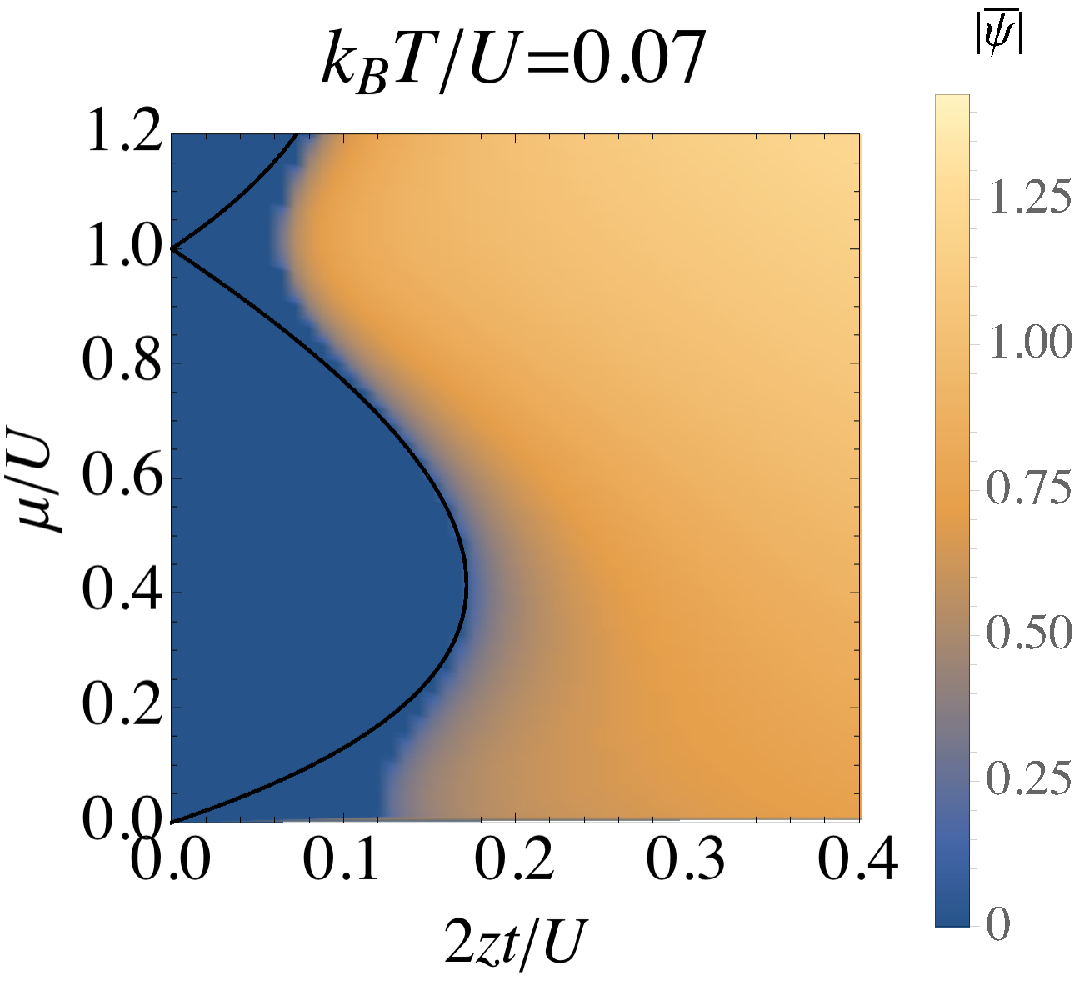}\includegraphics[width=1.7in]{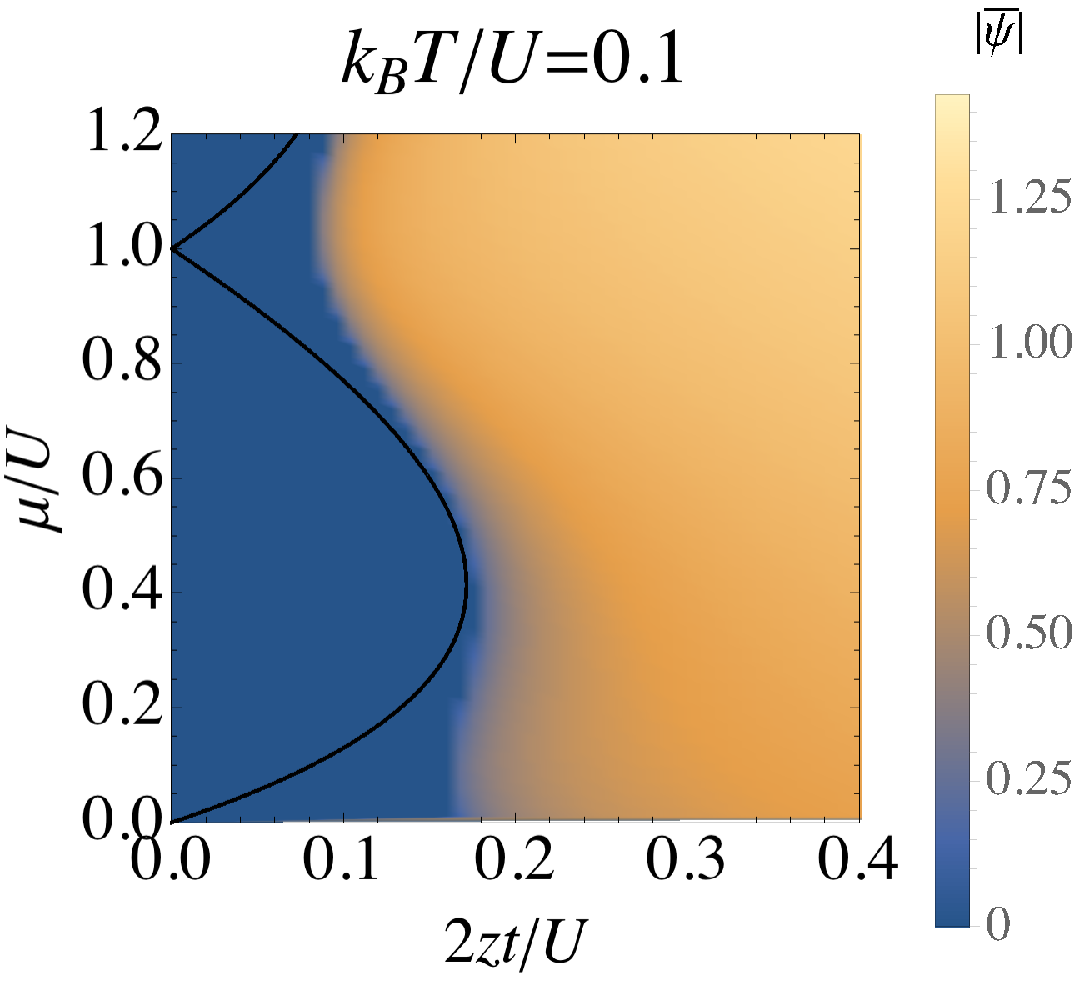}

\caption{Superfluid order parameter $|\bar{\psi}|$ as a function of the chemical
potential $\mu$ and hopping amplitude $t$ at different temperatures.
The calculations are performed within the homogeneous finite temperature
mean-field theory. The black curve in each plot corresponds to the
zero temperature Mott insulator-superfluid transition boundary obtained
by the mean-field theory employed in Refs.~\citep{Fisher_PRB_1989,Oosten_PRA_2001,Sachdev_QPT_2011}.
At nearly zero temperature ($k_{B}T/U=10^{-2}$), the distribution
of $|\bar{\psi}|$ on the $\mu$-$t$ plane still manifests a clear
Mott-lobe structure. As the temperature $T$ increases the superfluid
region with relatively small hopping amplitude $t$ vanishes first,
making the Mott-lobe structure less and less apparent. See text for
more details.}
\label{Fig_2_Mott_lobes_at_different_T_Hom_MFT}
\end{figure}

Moreover, away from unit filling, we calculated the superfluid order
parameter $|\bar{\psi}|$ as a function of the chemical potential
$\mu$ and hopping amplitude $t$ at different temperatures as shown
in Fig.~\ref{Fig_2_Mott_lobes_at_different_T_Hom_MFT}. We see that
at nearly zero temperature ($k_{B}T/U=10^{-2}$), the distribution
of $|\bar{\psi}|$ on the $\mu$-$t$ plane still manifests a clear
Mott-lobe structure, which agrees well with zero temperature results
from other mean-field type theories \citep{Fisher_PRB_1989,Jaksch_PRL_1998,Oosten_PRA_2001,Sachdev_QPT_2011}
(cf.~ the black curve in each plot of Fig.~\ref{Fig_2_Mott_lobes_at_different_T_Hom_MFT},
which corresponds to the zero temperature Mott insulator-superfluid
transition boundary obtained by the mean-field theory employed in
Refs.~\citep{Fisher_PRB_1989,Oosten_PRA_2001,Sachdev_QPT_2011}).
As $T$ increases the superfluid region with relatively small hopping
amplitude $t$ vanishes first, making the Mott-lobe structure less
and less apparent.

Finally, we remark that by comparing Eq.~(\ref{eq:HMF_homogeneous_ansatz})
with Eq.~(\ref{eq:H_SMFD_homogeneous}), we notice that under the
homogenous ansatz for $\psi_{\mathbf{i}}$, the mean-field Hamiltonian
constructed by the straightforward mean-field decoupling assumes exactly
the same form as the one obtained via the systematic functional integral
construction. On the one hand, this provides a solid theoretical ground
for the straightforward mean-field decoupling approach combined with
the homogenous ansatz. On the other hand, we should emphasize that
such a formal consistency is purely due to the coincidence caused
by the form of the homogeneous ansatz, under which only the negative
definite part of the quadrature associated with the hopping term {[}cf.~Eq.~(\ref{eq:qudrature_of_hopping_positive_negative_separation}){]}
eventually appears in the mean-field Hamiltonian. Indeed, as we shall
see explicitly in the following concrete example with an inhomogeneous
two-sublattice ansatz, the functional integral construction gives
rise to a mean-field Hamiltonian with a lower energy bound with respect
to the superfluid order parameter field, which is in sharp contrast
to the one constructed by the straightforward mean-field decoupling
shown in Eq.~(\ref{eq:H_SMFD_two_sub_lat}). 

\subsubsection{Two-sublattice finite temperature mean-field theory\label{subsec:Two-sublattice-FTMFT}}

Now, let us proceed to construct the mean-field theory by using a
more general ansatz for $\psi_{\mathbf{i}}$, which assumes a two-sublattice
structure, i.e., $\psi_{\mathbf{i}}=\psi_{\sigma}$ if $\mathbf{i}\in\mathring{\sigma}$
with $\sigma=e,o$, and $\mathring{\sigma}$ denoting the set of all
lattice sites of the sub-lattice with the index $\sigma$. Without
losing any generality, the explicit form of $\psi_{\mathbf{i}}$ under
this ansatz reads
\begin{equation}
\psi_{\mathbf{i}}=\frac{1+\cos(\mathbf{K}\cdot\mathbf{i})}{2}\psi_{e}+\frac{1-\cos(\mathbf{K}\cdot\mathbf{i})}{2}\psi_{o},\label{eq:two-sublattice_psi_explicit_form}
\end{equation}
with $\mathbf{K}\equiv(\pi,\pi)^{T}$. Directly plugging this ansatz
into Eq.~(\ref{eq:HMF_tight_binding_general_form}), we can directly
obtain (see Appendix \ref{App:Explicit_form_HMF} for derivation details)

\begin{align}
 & \hat{H}_{\mathrm{MF}}(\{\psi_{\sigma}^{*},\psi_{\sigma}\})\nonumber \\
= & N_{s}zt(\psi_{e}^{*}\psi_{e}+\psi_{o}^{*}\psi_{o})+\sum_{\mathbf{i}}\hat{H}_{\mathrm{SS}}^{(\mathbf{i})}(\{\psi_{\sigma}^{*},\psi_{\sigma}\}),\label{eq:HMF_two-sublattice_ansatz}
\end{align}
where
\begin{align}
 & \hat{H}_{\mathrm{SS}}^{(\mathbf{i})}(\{\psi_{\sigma}^{*},\psi_{\sigma}\})=\frac{1}{2}U\hat{n}_{\mathbf{i}}(\hat{n}_{\mathbf{i}}-1)-\mu\hat{n}_{\mathbf{i}}\label{eq:HSS_two-sublattice_ansatz}\\
 & +zt\left\{ \left[\hat{b}_{\mathbf{i}}^{\dagger}(\psi_{e}+\psi_{o})+\mathrm{h.c.}\right]+ie^{i\mathbf{K}\cdot\mathbf{i}}\left[\hat{b}_{\mathbf{i}}^{\dagger}(\psi_{e}-\psi_{o})+\mathrm{h.c.}\right]\right\} .\nonumber 
\end{align}
The corresponding mean-field partition function $Z_{\mathrm{MF}}$
reads 
\begin{align}
Z_{\mathrm{MF}} & =\int d(\{\psi_{\sigma}^{*},\psi_{\sigma}\})e^{-\beta N_{s}\Omega_{t,U,\mu,\beta}(\{\psi_{\sigma}^{*},\psi_{\sigma}\})},\label{eq:ZMF_two-sublattice_ansatz}
\end{align}
where 
\begin{align}
 & \Omega_{t,U,\mu,\beta}(\{\psi_{\sigma}^{*},\psi_{\sigma}\})\equiv zt\left(\psi_{e}^{*}\psi_{e}+\psi_{o}^{*}\psi_{o}\right)\\
 & -\frac{1}{2\beta}\ln\left(\mathrm{tr}\left[e^{-\beta\hat{H}_{\mathrm{SS}}^{(e)}(\{\psi_{\sigma}^{*},\psi_{\sigma}\})}\right]\cdot\mathrm{tr}\left[e^{-\beta\hat{H}_{\mathrm{SS}}^{(o)}(\{\psi_{\sigma}^{*},\psi_{\sigma}\})}\right]\right),\nonumber 
\end{align}
with $\hat{H}_{\mathrm{SS}}^{(\sigma)}(\psi_{\sigma}^{*},\psi_{\sigma})$
assuming the explicit form
\begin{align}
 & \hat{H}_{\mathrm{SS}}^{(\sigma)}(\{\psi_{\sigma}^{*},\psi_{\sigma}\})=\frac{1}{2}U\hat{n}(\hat{n}-1)-\mu\hat{n}\\
 & +zt\left\{ \left[\hat{b}^{\dagger}(\psi_{e}+\psi_{o})+\mathrm{h.c.}\right]+i\eta_{\sigma}\left[\hat{b}^{\dagger}(\psi_{e}-\psi_{o})+\mathrm{h.c.}\right]\right\} ,\nonumber 
\end{align}
where $\eta_{\sigma}\equiv+1,-1$ for $\sigma=e,o$, respectively.
From the explicit form of $\hat{H}_{\mathrm{SS}}^{(\sigma)}(\psi_{\sigma}^{*},\psi_{\sigma})$,
we notice that $\hat{H}_{\mathrm{SS}}^{(e)}$ is the hermitian conjugate
of $\hat{H}_{\mathrm{SS}}^{(o)}$, i.e., $(\hat{H}_{SS}^{(e)})^{\dagger}=\hat{H}_{SS}^{(o)}$.
This indicates the product $\mathrm{tr}[e^{-\beta\hat{H}_{SS}^{(e)}}]\cdot\mathrm{tr}[e^{-\beta\hat{H}_{SS}^{(o)}}]$
is guaranteed to assume positive real values. Therefore, the potential
function $\Omega_{t,U,\mu,\beta}(\{\psi_{\sigma}^{*},\psi_{\sigma}\})$
is still a real-valued function despite $\hat{H}_{\mathrm{MF}}(\{\psi_{\sigma}^{*},\psi_{\sigma}\})$
being a non-hermitian Hamiltonian. Moreover, following the same line
of derivation, one can also straightforwardly show that the potential
function $\Omega_{t,U,\mu,\beta}$ is real-valued for a class of ansatzes
for $\psi_{\mathbf{i}}$ assuming the form $\psi_{\mathbf{i}}=\varphi+\delta\varphi\cos(\mathbb{K}\cdot\mathbf{i})$
with $\mathbb{K}=(\pi/p,\pi/q)$ and $p(q)$ being a generic positive
integer, since $H_{\mathrm{SS}}^{(\mathbf{i})}(\{\psi_{\mathbf{i}}^{*},\psi_{\mathbf{i}}\})$
is essentially the hermitian conjugate of $H_{\mathrm{SS}}^{(\tilde{\mathbf{i}})}(\{\psi_{\mathbf{i}}^{*},\psi_{\mathbf{i}}\})$
with $\mathbb{K}\cdot(\mathbf{i}-\tilde{\mathbf{i}})=\pi$.

\begin{figure}
\includegraphics[width=1.7in]{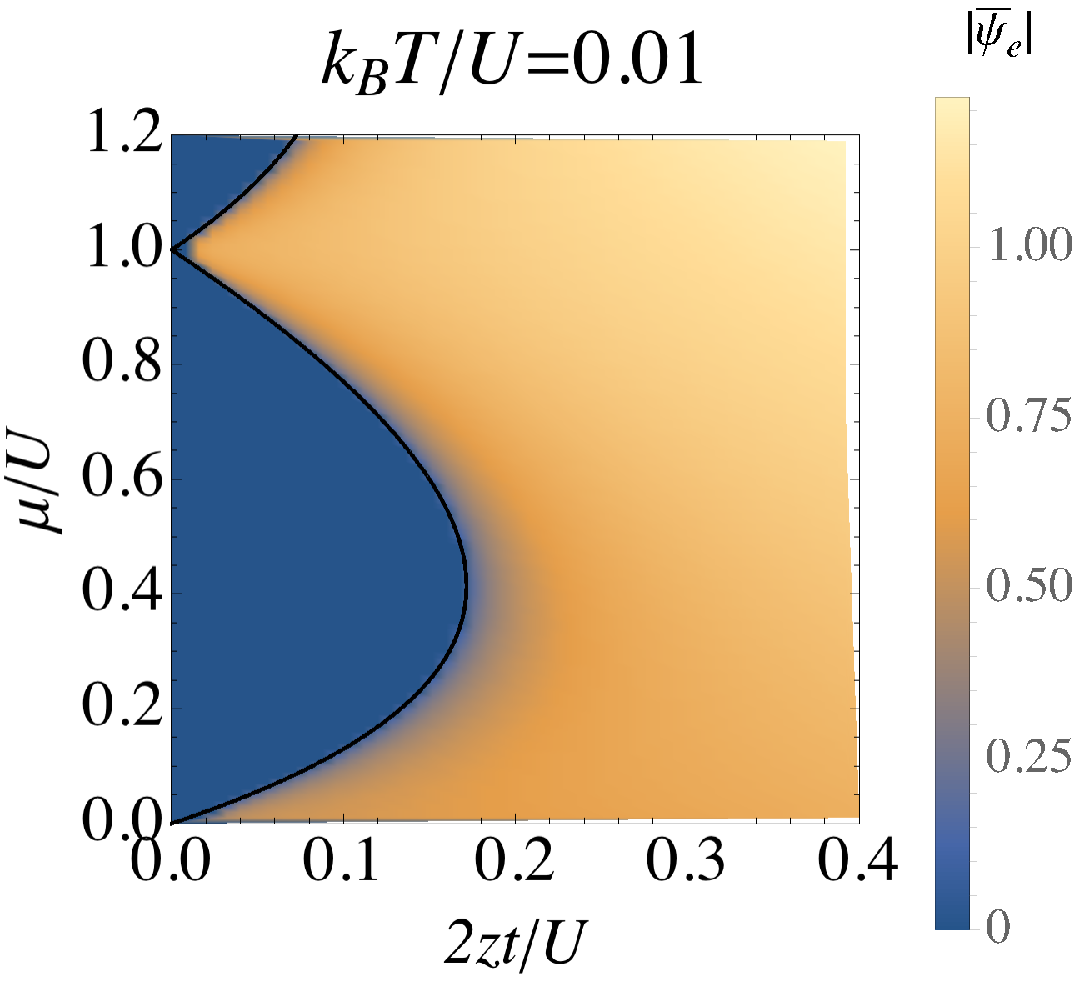}\includegraphics[width=1.7in]{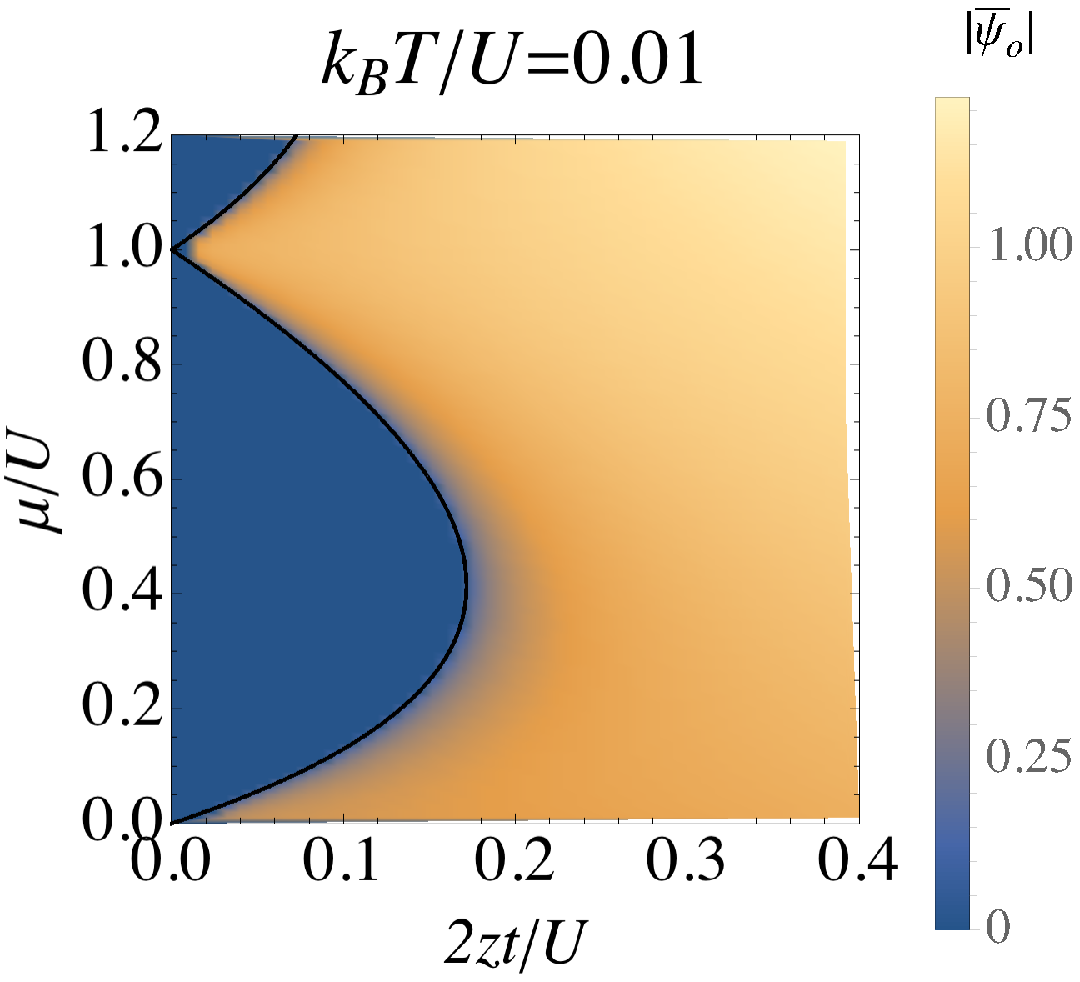}

\includegraphics[width=1.7in]{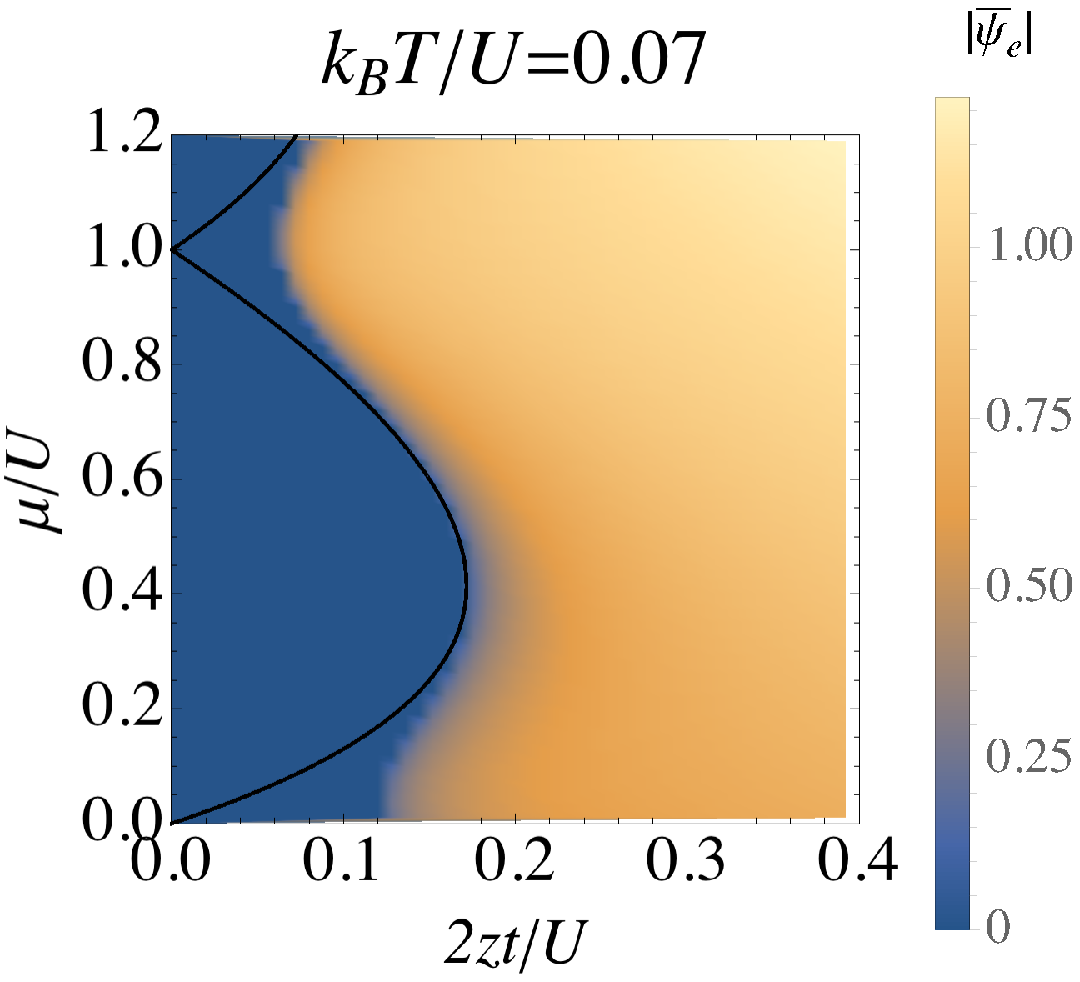}\includegraphics[width=1.7in]{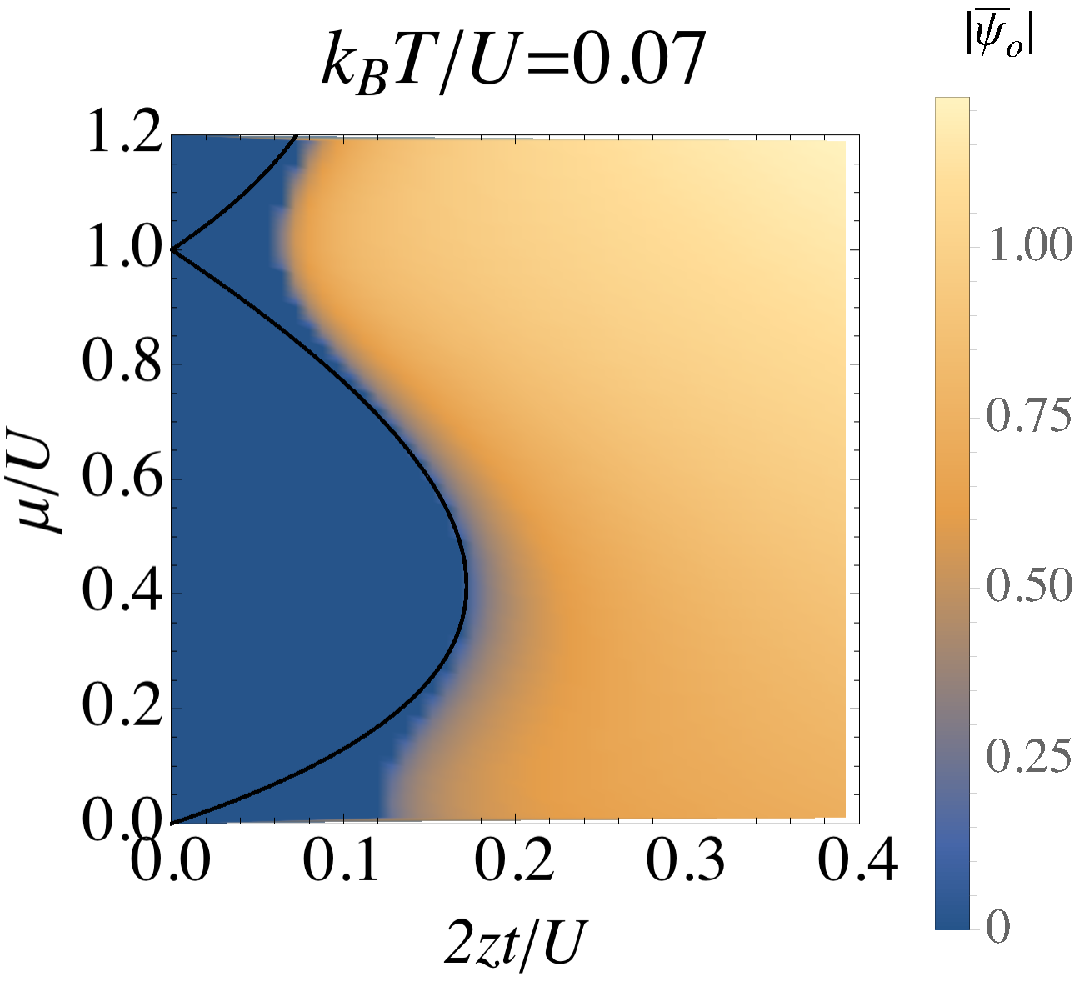}

\caption{Superfluid order parameter on the even and odd sublattice, i.e., $|\bar{\psi_{e}}|$
and $|\bar{\psi_{o}}|$, as a function of the chemical potential $\mu$
and hopping amplitude $t$ at different temperatures. The calculations
are performed within the two-sublattice finite temperature mean-field
theory. The black curve in each plot corresponds to the zero temperature
Mott insulator-superfluid transition boundary obtained by the mean-field
theory employed in Refs.~\citep{Fisher_PRB_1989,Oosten_PRA_2001,Sachdev_QPT_2011}.
The distribution of $|\bar{\psi}_{e}|$ on the $\mu$-$t$ plane is
exactly the same as the one of $|\bar{\psi}_{o}|$, indicating the
superfluid order parameter is always homogeneous over the whole system
as expected. Moreover, the $|\bar{\psi}_{e}|$ and the $|\bar{\psi}_{o}|$
distribution on the $\mu$-$t$ plane are exactly the same as the
$|\bar{\psi}|$ distribution obtained within the homogeneous mean-field
theory shown in Fig.~\ref{Fig_2_Mott_lobes_at_different_T_Hom_MFT}.
See text for more details.}
\label{Fig_3_Mott_lobes_at_different_T_two_sublattice_MFT}
\end{figure}

The concrete calculations based on the two-sublattice finite temperature
mean-field theory Eq.~(\ref{eq:ZMF_two-sublattice_ansatz}) can proceed
in a similar way as the one for the homogeneous mean-field theory.
At different temperatures, the superfluid order parameter on the even
and odd sublattice, i.e., $|\bar{\psi_{e}}|$ and $|\bar{\psi_{o}}|$,
as a function of the chemical potential $\mu$ and hopping amplitude
$t$ are shown in Fig.~\ref{Fig_3_Mott_lobes_at_different_T_two_sublattice_MFT}.
We directly observe that $|\bar{\psi_{e}}|$ and $|\bar{\psi_{o}}|$
show exactly the same behavior, indicating the superfluid order parameter
being homogeneous over the whole system as expected. Moreover, compared
with results from the homogeneous finite temperature mean-field theory
shown in Fig.~\ref{Fig_2_Mott_lobes_at_different_T_Hom_MFT}, we
can easily see that $|\bar{\psi_{e}}|$ and $|\bar{\psi_{o}}|$ manifest
the same behavior as $|\bar{\psi}|$, indicating the two-sublattice
finite temperature mean-field theory is consistent with the homogeneous
one. This is also consistent with the natural physical expectation
that equilibrium phases of the system should be homogeneous. 

\section{Conclusions }

The proper mean-field Hamiltonian can generally manifest distinct
intrinsic structure from its original one, as the finite temperature
mean-field theory for the Bose gases in optical lattices shows: due
to the general indefiniteness of the hopping matrix, the proper mean-field
treatment of the hopping term directly gives rise to a mean-field
Hamiltonian that is non-hermitian. This non-hermitian structure poses
no hindrance to the subsequent calculations. On the contrary, it facilitates
the application of the mean-field theory in combination with generic
space-dependent ansatzes for the order parameter field, based on which
an efficient and versatile approach for calculating finite temperature
properties of the system can be developed, as illustrated in the investigation
of the finite temperature superfluid transition of Bose gases in optical
lattices. As a direct application, our approach can be employed to
study finite temperature properties of lattice Bose gases in optical
cavities \citep{Landig_Nature_2016} whose order parameters can be
inhomogeneous in certain system parameter region. Also, we believe
our work will stimulate further efforts to investigate finite temperature
properties of the unconventional superfluids in ultracold atom systems
with spin-orbit coupling, effective magnetic flux, etc, and also other
distinct intrinsic structures of the mean-field Hamiltonians that
could appear in various mean-field treatments. Moreover, it is also
intriguing to promote the finite-temperature mean-field theory developed
here to systematically include corrections from quantum fluctuations
similar to what has been done in the zero-temperature case \citep{Santos_PRA_2009}.
\begin{acknowledgments}
We thank Lijun Lang for helpful discussions. This work was supported
by NSFC (Grant No.~11874017, No.~11674334, and No.~11947302), GDSTC
under Grant No.~2018A030313853, Science and Technology Program of
Guangzhou (Grant No.~2019050001), and START grant of South China
Normal University.
\end{acknowledgments}

\appendix

\section{Transformation from functional integral formulation back to operator
formulation\label{sec:Transformation-from-functional-back-to-operator}}

Using the expression of the mean-field partition function $Z_{\mathrm{MF}}$
shown in Eq.~(\ref{eq:mean-field_partition_function}), we can reformulate
it as 
\begin{align}
Z_{\mathrm{MF}} & =\int d(\Psi^{\dagger},\Psi)e^{-\beta\Psi^{\dagger}\mathbf{T}_{\mathrm{H}}\Psi}\nonumber \\
 & \times\int\mathcal{D}(B^{\dagger}(\tau),B(\tau))e^{-\tilde{S}_{(\Psi^{\dagger},\Psi)}[B^{\dagger}(\tau),B(\tau)]},
\end{align}
where 
\begin{align}
 & \tilde{S}_{(\Psi^{\dagger},\Psi)}[B^{\dagger}(\tau),B(\tau)]\equiv\nonumber \\
 & \int_{0}^{\beta}d\tau\sum_{\mathbf{i}}\left\{ b_{\mathbf{i}}^{*}\partial_{\tau}b_{\mathbf{i}}+\frac{U}{2}b_{\mathbf{i}}^{*}b_{\mathbf{i}}^{*}b_{\mathbf{i}}b_{\mathbf{i}}-\mu b_{\mathbf{i}}^{*}b_{\mathbf{i}}\right.\nonumber \\
 & \left.-\left[\left(\Psi^{\dagger}\mathbf{T}_{\mathrm{NH}}\right)_{\mathbf{i}}b_{\mathbf{i}}+b_{\mathbf{i}}^{*}\left(\mathbf{T}_{\mathrm{NH}}\Psi\right)_{\mathbf{i}}\right]\right\} .\label{eq:S_tilde}
\end{align}
Here, we notice that since $\Psi$ is $\tau$-independent in the mean-field
approximation, $\tilde{S}_{(\Psi^{\dagger},\Psi)}[B^{\dagger}(\tau),B(\tau)]$
only assumes a parameter dependence on $(\Psi^{\dagger},\Psi)$, hence
it is a functional of $(B^{\dagger}(\tau),B(\tau))$ only. This indicates
that the functional integral with respect to $(B^{\dagger}(\tau),B(\tau))$
with $e^{-\tilde{S}_{(\Psi^{\dagger},\Psi)}[B^{\dagger}(\tau),B(\tau)]}$
being the integrand directly corresponds to a partition function of
an effective Hamiltonian $\hat{\tilde{H}}(\Psi^{\dagger},\Psi)$ which
assumes a parameter dependence on $(\Psi^{\dagger},\Psi)$, i.e.,
\begin{equation}
\int\mathcal{D}(B^{\dagger}(\tau),B(\tau))e^{-\tilde{S}_{(\Psi^{\dagger},\Psi)}[B^{\dagger}(\tau),B(\tau)]}=\mathrm{tr}\left[e^{-\beta\hat{\tilde{H}}(\Psi^{\dagger},\Psi)}\right].
\end{equation}
Using the standard coherent state functional integral approach, one
can straightforwardly show that the Hamiltonian $\hat{\tilde{H}}(\Psi^{\dagger},\Psi)$
that satisfies the above equation assumes the explicit form 
\begin{align}
\hat{\tilde{H}}(\Psi^{\dagger},\Psi)= & \sum_{\mathbf{i}}\frac{U}{2}\hat{b}_{\mathbf{i}}^{\dagger}\hat{b}_{\mathbf{i}}^{\dagger}\hat{b}_{\mathbf{i}}\hat{b}_{\mathbf{i}}-\mu\hat{b}_{\mathbf{i}}^{\dagger}\hat{b}_{\mathbf{i}}\nonumber \\
 & -\left[\left(\Psi^{\dagger}\mathbf{T}_{\mathrm{NH}}\right)_{\mathbf{i}}\hat{b}_{\mathbf{i}}+\hat{b}_{\mathbf{i}}^{\dagger}\left(\mathbf{T}_{\mathrm{NH}}\Psi\right)_{\mathbf{i}}\right].
\end{align}
Therefore, we can rewrite $Z_{\mathrm{MF}}$ as the form shown in
Eq.~(\ref{eq:Z_MF}), i.e.,
\begin{align}
Z_{\mathrm{MF}} & =\int d(\Psi^{\dagger},\Psi)e^{-\beta\Psi^{\dagger}T_{\mathrm{H}}\Psi}\mathrm{tr}\left[e^{-\beta\hat{\tilde{H}}(\Psi^{\dagger},\Psi)}\right]\nonumber \\
 & =\int d(\Psi^{\dagger},\Psi)\mathrm{tr}\left[e^{-\beta\hat{H}_{\mathrm{MF}}(\Psi^{\dagger},\Psi)}\right],
\end{align}
with $\hat{H}_{\mathrm{MF}}(\Psi^{\dagger},\Psi)=\Psi^{\dagger}\mathbf{T}_{\mathrm{H}}\Psi+\hat{\tilde{H}}(\Psi^{\dagger},\Psi)$.

\section{Explicit form of the mean-field Hamiltonian under the homogeneous
and the inhomogeneous two-sub-lattice ansatz\label{App:Explicit_form_HMF}}

In this appendix, we present the relevant derivation details for the
explicit form of $\hat{H}_{\mathrm{MF}}(\Psi^{\dagger},\Psi)$ in
the general case, and also its form under the homogeneous and the
inhomogeneous two-sub-lattice ansatz for $\psi_{\mathbf{i}}$.

\subsection{Explicit form of $\hat{H}_{\mathrm{MF}}(\Psi^{\dagger},\Psi)$}

From Eq.~(\ref{eq:hopping_term_diagonal_form}) we see that the hopping
term is directly diagonalized by the lattice Fourier transformation.
With the explicit form of the lattice Fourier transformation, one
can directly write down the following relations concerning the corresponding
unitary matrix $U$ that diagonalizes the hopping matrix $T$, i.e.,
\begin{align}
U(\cdots,\hat{b}_{\mathbf{i}},\cdots)^{T} & =(\cdots,\sum_{\mathbf{j}}\frac{e^{-i\mathbf{k}\cdot\mathbf{j}}}{\sqrt{N_{s}}}\hat{b}_{\mathbf{j}},\cdots)^{T},\\
(\cdots,\hat{b}_{\mathbf{i}}^{\dagger},\cdots)U^{\dagger} & =(\cdots,\sum_{\mathbf{j}}\frac{e^{i\mathbf{k}\cdot\mathbf{j}}}{\sqrt{N_{s}}}\hat{b}_{\mathbf{j}}^{\dagger},\cdots),\\
U\Psi & =(\cdots,\sum_{\mathbf{i}}\frac{e^{-i\mathbf{k}\cdot\mathbf{i}}}{\sqrt{N_{s}}}\psi_{\mathbf{i}},\cdots)^{T},\\
\Psi^{\dagger}U^{\dagger} & =(\cdots,\sum_{\mathbf{i}}\frac{e^{i\mathbf{k}\cdot\mathbf{i}}}{\sqrt{N_{s}}}\psi_{\mathbf{i}}^{*},\cdots).
\end{align}
From Eq.~(\ref{eq:HMF_general_form}) we see that in order to obtain
the explicit form of $\hat{H}_{\mathrm{MF}}(\Psi^{\dagger},\Psi)$,
one needs to get the explicit form for the quadratures, $\Psi^{\dagger}\mathbf{T}_{\mathrm{H}}\Psi$,
$\sum_{\mathbf{i}}(\Psi^{\dagger}\mathbf{T}_{\mathrm{NH}})_{\mathbf{i}}\hat{b}_{\mathbf{i}}$,
and $\sum_{\mathbf{i}}\hat{b}_{\mathbf{i}}^{\dagger}(\mathbf{T}_{\mathrm{NH}}\Psi)_{\mathbf{i}}$.
This can be done by plugging the explicit forms of $\mathbf{T}_{\mathrm{H}}$
and $\mathbf{T}_{\mathrm{NH}}$ in Eqs.~(\ref{eq:T_H}, \ref{eq:T_NH}),
and the above relations for the unitary matrix $U$ into the explicit
form of the quadratures, i.e.
\begin{align}
 & \Psi^{\dagger}\mathbf{T}_{\mathrm{H}}\Psi=\Psi^{\dagger}U^{\dagger}\left(E_{(+)}\bigoplus(-E_{(-)})\right)U\Psi\nonumber \\
 & =\sum_{\mathbf{k},\mathbf{i},\mathbf{i}'}\frac{e^{i\mathbf{k}\cdot\mathbf{i}}}{\sqrt{N_{s}}}\psi_{\mathbf{i}}^{*}[\theta(\mathcal{E}(\mathbf{k}))-\theta(-\mathcal{E}(\mathbf{k}))]\mathcal{E}(\mathbf{k})\frac{e^{-i\mathbf{k}\cdot\mathbf{i}'}}{\sqrt{N_{s}}}\psi_{\mathbf{i}'}\nonumber \\
 & =\frac{1}{N_{s}}\sum_{\mathbf{i},\mathbf{i}',\mathbf{k}}\psi_{\mathbf{i}}^{*}\left([\theta(\mathcal{E}(\mathbf{k}))-\theta(-\mathcal{E}(\mathbf{k}))]\mathcal{E}(\mathbf{k})e^{i\mathbf{k}\cdot(\mathbf{i}-\mathbf{i}')}\right)\psi_{\mathbf{i}'},\label{eq:quadrature_1_explicit_form}
\end{align}
\begin{align}
 & \sum_{\mathbf{i}}(\Psi^{\dagger}\mathbf{T}_{\mathrm{NH}})_{\mathbf{i}}\hat{b}_{\mathbf{i}}=\sum_{\mathbf{i}}\left(\Psi^{\dagger}U^{\dagger}\left(iE_{(+)}\bigoplus(E_{(-)})\right)U\right)_{\mathbf{i}}\hat{b}_{\mathbf{i}}\nonumber \\
 & =\sum_{\mathbf{k},\mathbf{i},\mathbf{i}'}\frac{e^{i\mathbf{k}\cdot\mathbf{i}}}{\sqrt{N_{s}}}\psi_{\mathbf{i}}^{*}[\theta(-\mathcal{E}(\mathbf{k}))+i\theta(\mathcal{E}(\mathbf{k}))]\mathcal{E}(\mathbf{k})\frac{e^{-i\mathbf{k}\cdot\mathbf{i}'}}{\sqrt{N_{s}}}\hat{b}_{\mathbf{i}'}\nonumber \\
 & =\frac{1}{N_{s}}\sum_{\mathbf{i},\mathbf{i}',\mathbf{k}}\psi_{\mathbf{i}}^{*}\left([\theta(-\mathcal{E}(\mathbf{k}))+i\theta(\mathcal{E}(\mathbf{k}))]\mathcal{E}(\mathbf{k})e^{i\mathbf{k}\cdot(\mathbf{i}-\mathbf{i}')}\right)\hat{b}_{\mathbf{i}'},\label{eq:quadrature_2_explicit_form}
\end{align}
\begin{align}
 & \sum_{\mathbf{i}}\hat{b}_{\mathbf{i}}^{\dagger}(\mathbf{T}_{\mathrm{NH}}\Psi)_{\mathbf{i}}=\sum_{\mathbf{i}}\hat{b}_{\mathbf{i}}^{\dagger}\left(U^{\dagger}\left(iE_{(+)}\bigoplus(E_{(-)})\right)U\Psi\right)_{\mathbf{i}}\nonumber \\
 & =\sum_{\mathbf{k},\mathbf{i},\mathbf{i}'}\frac{e^{i\mathbf{k}\cdot\mathbf{i}}}{\sqrt{N_{s}}}\hat{b}_{\mathbf{i}}^{\dagger}[\theta(-\mathcal{E}(\mathbf{k}))+i\theta(\mathcal{E}(\mathbf{k}))]\mathcal{E}(\mathbf{k})\frac{e^{-i\mathbf{k}\cdot\mathbf{i}'}}{\sqrt{N_{s}}}\psi_{\mathbf{i}'}\nonumber \\
 & =\frac{1}{N_{s}}\sum_{\mathbf{i},\mathbf{i}',\mathbf{k}}\hat{b}_{\mathbf{i}}^{\dagger}\left([\theta(-\mathcal{E}(\mathbf{k}))+i\theta(\mathcal{E}(\mathbf{k}))]\mathcal{E}(\mathbf{k})e^{i\mathbf{k}\cdot(\mathbf{i}-\mathbf{i}')}\right)\psi_{\mathbf{i}'}.\label{eq:quadrature_3_explicit_form}
\end{align}
With the above explicit forms of the quadratures, we directly can
get the explicit form for $\hat{H}_{\mathrm{MF}}(\Psi^{\dagger},\Psi)$
shown in Eq.~(\ref{eq:HMF_tight_binding_general_form}).

\subsection{Explicit form of $\hat{H}_{\mathrm{MF}}(\psi^{*},\psi)$ in the homogeneous
finite temperature mean-field theory}

To obtain the explicit form of $\hat{H}_{\mathrm{MF}}(\psi^{*},\psi)$,
we directly plug the homogeneous ansatz $\psi_{\mathbf{i}}=\psi$
into the explicit forms of the three quadratures shown in Eqs.~(\ref{eq:quadrature_1_explicit_form},
\ref{eq:quadrature_2_explicit_form}, \ref{eq:quadrature_3_explicit_form}),
and can obtain
\begin{align}
 & \Psi^{\dagger}\mathbf{T}_{\mathrm{H}}\Psi\nonumber \\
= & \frac{1}{N_{s}}\sum_{\mathbf{i},\mathbf{i}',\mathbf{k}}\psi^{*}\left([\theta(\mathcal{E}(\mathbf{k}))-\theta(-\mathcal{E}(\mathbf{k}))]\mathcal{E}(\mathbf{k})e^{i\mathbf{k}\cdot(\mathbf{i}-\mathbf{i}')}\right)\psi\nonumber \\
= & 2ztN_{s}\psi^{*}\psi,
\end{align}
\begin{align}
 & \sum_{\mathbf{i}}(\Psi^{\dagger}\mathbf{T}_{\mathrm{NH}})_{\mathbf{i}}\hat{b}_{\mathbf{i}}\nonumber \\
= & \frac{1}{N_{s}}\sum_{\mathbf{i},\mathbf{i}',\mathbf{k}}\psi^{*}\left([\theta(-\mathcal{E}(\mathbf{k}))+i\theta(\mathcal{E}(\mathbf{k}))]\mathcal{E}(\mathbf{k})e^{i\mathbf{k}\cdot(\mathbf{i}-\mathbf{i}')}\right)\hat{b}_{\mathbf{i}'}\nonumber \\
= & -2zt\sum_{\mathbf{i}}\psi^{*}\hat{b}_{\mathbf{i}},
\end{align}
\begin{align}
 & \sum_{\mathbf{i}}\hat{b}_{\mathbf{i}}^{\dagger}(\mathbf{T}_{\mathrm{NH}}\Psi)_{\mathbf{i}}\nonumber \\
= & \frac{1}{N_{s}}\sum_{\mathbf{i},\mathbf{i}',\mathbf{k}}\hat{b}_{\mathbf{i}}^{\dagger}\left([\theta(-\mathcal{E}(\mathbf{k}))+i\theta(\mathcal{E}(\mathbf{k}))]\mathcal{E}(\mathbf{k})e^{i\mathbf{k}\cdot(\mathbf{i}-\mathbf{i}')}\right)\psi\nonumber \\
= & -2zt\sum_{\mathbf{i}}\hat{b}_{\mathbf{i}}^{\dagger}\psi,
\end{align}
where the explicit expression for $\mathcal{E}(\mathbf{k})$, i.e.,
$\mathcal{E}(\mathbf{k})=-4t\left(\cos k_{x}+\cos k_{y}\right)$,
is also used in the above derivations. With the above explicit forms
of the quadratures, we directly get the explicit form of $\hat{H}_{\mathrm{MF}}(\psi^{*},\psi)$
shown in Eq.~(\ref{eq:HMF_homogeneous_ansatz}).

\subsection{Explicit form of $\hat{H}_{\mathrm{MF}}(\{\psi_{\sigma}^{*},\psi_{\sigma}\})$
in the inhomogeneous two-sublattice finite temperature mean-field
theory}

To obtain the explicit form of $\hat{H}_{\mathrm{MF}}(\{\psi_{\sigma}^{*},\psi_{\sigma}\})$,
we first rewrite two-sublattice ansatz for $\psi_{\mathbf{i}}$ shown
in Eq.~(\ref{eq:two-sublattice_psi_explicit_form}) into a more convenient
form for calculations, i.e., 
\begin{equation}
\psi_{\mathbf{i}}=\varphi+\delta\varphi\frac{e^{i\mathbf{K}\cdot\mathbf{i}}+e^{-i\mathbf{K}\cdot\mathbf{i}}}{2},\label{eq:psi_i_convenient_form}
\end{equation}
where 
\begin{equation}
\varphi\equiv\frac{\psi_{e}+\psi_{o}}{2},\delta\varphi\equiv\frac{\psi_{e}-\psi_{o}}{2}.
\end{equation}
We then directly plug Eq.~(\ref{eq:psi_i_convenient_form}) into
the explicit forms of the three quadratures shown in Eqs.~(\ref{eq:quadrature_1_explicit_form},
\ref{eq:quadrature_2_explicit_form}, \ref{eq:quadrature_3_explicit_form}),
and can obtain
\begin{align}
 & \Psi^{\dagger}\mathbf{T}_{\mathrm{H}}\Psi=\frac{1}{N_{s}}\sum_{\mathbf{i},\mathbf{i}',\mathbf{k}}\left(\varphi^{*}+\delta\varphi^{*}\frac{e^{i\mathbf{K}\cdot\mathbf{i}}+e^{-i\mathbf{K}\cdot\mathbf{i}}}{2}\right)\nonumber \\
 & \times\left([\theta(\mathcal{E}(\mathbf{k}))-\theta(-\mathcal{E}(\mathbf{k}))]\mathcal{E}(\mathbf{k})e^{i\mathbf{k}\cdot(\mathbf{i}-\mathbf{i}')}\right)\nonumber \\
 & \times\left(\varphi+\delta\varphi\frac{e^{i\mathbf{K}\cdot\mathbf{i}'}+e^{-i\mathbf{K}\cdot\mathbf{i}'}}{2}\right)\nonumber \\
= & 8tN_{s}(\varphi^{*}\varphi+\delta\varphi^{*}\delta\varphi)\nonumber \\
= & ztN_{s}(\psi_{e}^{*}\psi_{e}+\psi_{o}^{*}\psi_{o}),\label{eq:qudrature_1_two-sublattice_ansatz}
\end{align}
\begin{align}
 & \sum_{\mathbf{i}}(\Psi^{\dagger}\mathbf{T}_{\mathrm{NH}})_{\mathbf{i}}\hat{b}_{\mathbf{i}}=\frac{1}{N_{s}}\sum_{\mathbf{i},\mathbf{i}',\mathbf{k}}\left(\varphi^{*}+\delta\varphi^{*}\frac{e^{i\mathbf{K}\cdot\mathbf{i}}+e^{-i\mathbf{K}\cdot\mathbf{i}}}{2}\right)\nonumber \\
 & \times\left([\theta(-\mathcal{E}(\mathbf{k}))+i\theta(\mathcal{E}(\mathbf{k}))]\mathcal{E}(\mathbf{k})e^{i\mathbf{k}\cdot(\mathbf{i}-\mathbf{i}')}\right)\hat{b}_{\mathbf{i}'},\nonumber \\
 & =-8t\sum_{\mathbf{i}}\hat{b}_{\mathbf{i}}\left(\varphi^{*}+ie^{i\mathbf{K}\cdot\mathbf{i}}\delta\varphi^{*}\right)\nonumber \\
 & =-zt\sum_{\mathbf{i}}\hat{b}_{\mathbf{i}}\left[(\psi_{e}^{*}+\psi_{o}^{*})+ie^{i\mathbf{K}\cdot\mathbf{i}}(\psi_{e}^{*}-\psi_{o}^{*})\right],\label{eq:qudrature_2_two-sublattice_ansatz}
\end{align}
\begin{align}
 & \sum_{\mathbf{i}}\hat{b}_{\mathbf{i}}^{\dagger}(\mathbf{T}_{\mathrm{NH}}\Psi)_{\mathbf{i}}=\frac{1}{N_{s}}\sum_{\mathbf{i},\mathbf{i}',\mathbf{k}}\left(\varphi+\delta\varphi\frac{e^{i\mathbf{K}\cdot\mathbf{i}'}+e^{-i\mathbf{K}\cdot\mathbf{i}'}}{2}\right)\nonumber \\
 & \times\hat{b}_{\mathbf{i}}^{\dagger}\left([\theta(-\mathcal{E}(\mathbf{k}))+i\theta(\mathcal{E}(\mathbf{k}))]\mathcal{E}(\mathbf{k})e^{i\mathbf{k}\cdot(\mathbf{i}-\mathbf{i}')}\right)\nonumber \\
 & =-8t\sum_{\mathbf{i}}\hat{b}_{\mathbf{i}}^{\dagger}\left(\varphi+ie^{i\mathbf{K}\cdot\mathbf{i}}\delta\varphi\right)\nonumber \\
 & =-zt\sum_{\mathbf{i}}\hat{b}_{\mathbf{i}}^{\dagger}\left[(\psi_{e}+\psi_{o})+ie^{i\mathbf{K}\cdot\mathbf{i}}(\psi_{e}-\psi_{o})\right],\label{eq:qudrature_3_two-sublattice_ansatz}
\end{align}
where the explicit expression for $\mathcal{E}(\mathbf{k})$, i.e.,
$\mathcal{E}(\mathbf{k})=-4t\left(\cos k_{x}+\cos k_{y}\right)$,
is also used in the above derivations. With the above explicit forms
of the quadratures, we can directly get the explicit form of $\hat{H}_{\mathrm{MF}}(\{\psi_{\sigma}^{*},\psi_{\sigma}\})$
shown in Eq.~(\ref{eq:HMF_two-sublattice_ansatz}).

\end{document}